\documentstyle[12pt,psfig]{article}
\begin{document}

\vskip 1truecm
\rightline{Preprint PUP-TH-1618}
\rightline{ e-Print Archive: hep-lat/9605001}
\bigskip
\centerline{\Large  Improved Hamiltonian for Minkowski Yang-Mills Theory}
\bigskip
\centerline{\Large Guy D. Moore\footnote{e-mail:
guymoore@puhep1.princeton.edu}
}
\medskip

\centerline{\it Princeton University}
\centerline{\it Joseph Henry Laboratories, PO Box 708}
\centerline{\it Princeton, NJ 08544, USA}

\medskip

\centerline{\bf Abstract}

\smallskip

I develop an improved Hamiltonian for classical, Minkowski 
Yang-Mills theory, which evolves
infrared fields with corrections from lattice spacing $a$ beginning at
$O(a^4)$.  I use it to investigate the response of Chern-Simons number 
to a chemical potential, and to compute the maximal Lyapunov exponent.  
Both quantities have small $a$ limits, in both cases within $10\% $
of the limit found using the unimproved (Kogut Susskind) Hamiltonian.
For the maximal Lyapunov exponent the limits 
differ by about $5 \% $, significant
at about $5 \sigma$, indicating that while a small $a$ limit exists,
its value is corrupted by lattice artefacts.  
For the response of Chern-Simons number the
statistics are not good enough to resolve $ 5 \% $ differences, but it
seems possible in analogy with the Lyapunov exponent that the final answer
depends on the lattice regulation.

\section{Introduction}

Several interesting and outstanding questions in high energy physics and
cosmology, such as the production of quark-gluon plasma in relativistic
heavy ion collisions and the conditions in the early universe under which
baryogenesis may have occurred, involve finite temperature field theories
where there are interactions between bosons, and some of the bosons have
very small masses in comparison to the temperature, $m \ll T$.
The infrared physics of such systems  can be strongly
coupled; in perturbation theory, the contribution of loops involving 
very soft ($O(g^2 T)$, $g$ a generic coupling strength characterizing the
theory) momenta can be order 1, 
even when $g$ is small enough that the vacuum theory
is weakly coupled.  This strongly coupled infrared
physics is often very interesting; in the electroweak theory it may
determine
the strength of the electroweak phase transition, and it sets the rate
of Chern-Simons number ($N_{CS}$) diffusion, which determines the rate of
baryon number violation.  The former problem is thermodynamical, and there
are Euclidean tools which can be brought to bear on the nonperturbative
physics involved\cite{FKLRS,Fodor}.  However, the latter question is 
dynamical, and cannot be addressed by Euclidean methods.  There is also
interest in a more general understanding of the infrared dynamics of Yang-Mills
and Yang-Mills Higgs theory at finite temperature, for instance to 
understand its implications in heavy ion collisions.  These questions
cannot be addressed by Euclidean methods, which are only appropriate for
answering thermodynamical questions.  They are also very hard to address
analytically in any reliable way, and the prospects for numerical evaluation
of Minkowski path integrals are more or less hopeless.

Several years ago, Grigoriev and Rubakov made an interesting proposal for
a new numerical approach to the understanding of the infrared problem
in finite temperature bosonic systems\cite{GrigorievRubakov}.  The key
observation is that, if the coupling constant $g$ is small, then the
nonperturbative physics arises because occupation numbers are large in the
infrared; but this is precisely the regime where a classical
field approximation becomes reasonable.  The physics is 
nonperturbative but in an essentially classical way.  Therefore, it is 
reasonable to hope that, by analyzing the analogous classical field
theory at finite temperature, one can understand at leading order the 
physics of the quantum system of interest.

Based on this idea, there has been a revival of interest in classical,
thermal Yang-Mills 
theory\cite{Ambjornetal,AmbjornFarakos,Lyapunovguys,Krasnitz,AmbKras,paper1}.
The work is based on lattice implementation of classical Yang-Mills theory,
using the Hamiltonian formulism worked out by Kogut and Susskind years
ago\cite{KogutSusskind}.  This work has yielded interesting results;
it appears that we now know the $N_{CS}$ diffusion rate in the symmetric
electroweak phase to within around $5\%$ accuracy\cite{AmbKras},
and there is numerical evidence for the interesting conjecture that the
maximal Lyapunov exponent of Yang-Mills theory is twice the damping
rate of a plasmon at rest\cite{Lyapunovguys}.

However, there have been some frustrations in the application of this
technology; for instance, in \cite{paper1}
I find that, to determine the rate of $N_{CS}$ diffusion
in SU(3) gauge theory, the lattice spacing must be so fine that the
calculation is computationally prohibitive.  Also, there is a more 
fundamental challenge to the whole approach, raised by
Bodecker et. al. \cite{McLerranetal},
who consider the ``hard thermal loops,'' 
the one loop contributions of ultraviolet modes to the dispersion
relations and interactions of  
the infrared modes.  On the lattice, the strength of the
hard thermal loops grows linearly with inverse lattice spacing, and
as is often the case with linearly divergent quantities, the functional
form of the hard thermal loops may depend on the form of the cutoff; they 
posess lattice artefacts which make their $\vec{q}/q_0$
dependence different than in the continuum, quantum theory.
Although the results for $N_{CS}$ diffusion in SU(2) apparently have a
fine lattice limit\cite{AmbKras}, which must be independent 
of the overall amplitude of the hard thermal loops, 
the value of the fine lattice limit may depend on the functional form of the
hard thermal loops, and hence on the specifics of the lattice cutoff.  I
briefly explored this possibility in \cite{paper1}, where I found some
evidence for a weak dependence on the form of the cutoff; however the case
was far from clear, and it is an open and interesting question whether
the $N_{CS}$ diffusion rate depends on the presence of hard thermal
loops and on their functional form, and whether the convergence to a 
fine lattice limit represents the disappearance of nonrenormalizeable
operators or a suppression of effects which die as a power of 
$m_D^2 /m^2_{\rm nonperturbative}$.

One or both of these problems might be solved by finding an ``improved''
Hamiltonian for the classical, Minkowski Yang-Mills theory.  By ``improved''
I mean a Hamiltonian such that the interaction of infrared modes would
be the same as in the continuum, plus finite $a$ effects starting at
$a^4$, rather than at $a^2$, as they do in the simplest lattice 
implementation.  That is, for infrared fields the lattice action can
be expanded in operator dimension, and an ``improved'' action is one
in which, in addition to the desired dimension 4 $F_{\mu \nu} F^{\mu \nu}$
term, there are no dimension 6 operators, but only those starting at
dimension 8.  It is well known that in evolving scalar partial differential
equations, and in Euclidean lattice gauge theory, such improvement greatly
reduces the numerical demands by allowing accurate results on much
coarser lattices\cite{Lepage}.  Implementing such an improved Hamiltonian
for the systems under consideration should either lighten the numerical
demands, or verify that hard thermal loops are indeed important, either 
because the lattice fineness requirement arises from the need to make
the Debye mass large, or because the results depend on the functional 
form of the hard thermal loops, and hence on the form of the
regulation.

The basic idea of an improved lattice Hamiltonian is quite simple, and
is easiest to present in the case of scalar field theory.  Consider the 
evolution of the continuum system derived from the Lagrangian
\begin{equation}
L = \int d^3 x \left(  \frac{\dot{\phi}^2}{2} -
\frac{\vec{(\nabla} \phi)^2}{2} - \frac{m^2}{2} \phi^2 
- \frac{\lambda}{4} \phi^4  \right) \, ,
\end{equation}
with $\phi$ a single component, real scalar field.  The
Hamilton's equations of motion (defining $\pi$ as the conjugate momentum
of $\phi$) are
\begin{eqnarray}
\frac{d\phi}{dt} & = & \pi \, ,
\\
\frac{d \pi}{dt} & = &  \nabla^2 \phi + m^2 \phi + \lambda \phi^3 \, . 
\end{eqnarray}

To implement these equations on a lattice, we discretize the fields $\phi$
and $\pi$ so they each take on a single value at each lattice point.  The
only difficulty lies in choosing a lattice definition of $\nabla^2$.
If we make the replacement
\begin{equation}
\int d^3x (\vec{\nabla} \phi)^2 = a^3 \sum_x \sum_{\hat{i}} 
\frac{ (\phi(x + a \hat{i}) - \phi(x))^2}{a^2}
\end{equation}
in the Lagrangian, then the appropriate substitution in the equation
of motion is
\begin{equation}
\nabla^2 \phi(x) \rightarrow \sum_{\hat{i}} \frac{ \phi( x- a \hat{i})
+ \phi(x + a \hat{i}) - 2 \phi(x)}{a^2}
\end{equation}
which is the standard lattice Laplacian.

Now consider the quality of this implementation when evolving some smooth
field configuration.  If the characteristic wavelength of the
field configuation of interest is on order the lattice length, it is
inevitable that the lattice implementation will poorly reproduce the
continuum dynamics; so it is only interesting to test the quality of
the lattice implementation for fields which are smooth on the lattice
scale, in which case we can study the quality of the lattice Laplacian
by expanding $\phi$ in a Taylor series about the point $x$; it is 
straightforward to show that
\begin{equation}
\sum_{\hat{i}} \frac{ \phi( x- a \hat{i})
+ \phi(x + a \hat{i}) - 2 \phi(x)}{a^2} = \sum_i \nabla_i^2 \phi
+ \frac{a^2}{12} \sum_i \nabla_i^4 \phi
 + \frac{a^4}{360} \sum_i \nabla_i^6 \phi + \ldots
\end{equation}
where I write in terms of a positive 3 dimensional measure, as I will
for the remainder of the paper.  All but the first of the derivative
terms are undesirable.  The second term will produce lattice artefacts
in the evolution of a mode of wave vector $k$ which will be on order
$a^2 k^2$; if $a \ll 1/k$, which is necessary for the lattice
implementation to at all resemble the desired continuum system, then
the quality of the implementation would be improved if we could write
a new lattice Laplacian which did not produce an $O(a^2)$ term.

This is quite easy; just include contributions from next nearest neighbors
as well as nearest neighbors, and choose the coefficients so the
coefficients of $\nabla^2$ and $\nabla_i^4$ are as desired.  The correct
choice is
\begin{eqnarray}
\nabla^2_{\rm Latt} & = & \sum_i \frac{ \frac{4}{3} ( \phi ( x + a\hat{i} ) +
 \phi ( x - a\hat{i} ) ) - \frac{1}{12} ( \phi ( x + 2a\hat{i} )
+  \phi ( x - 2a\hat{i} )) - \frac{5}{2} \phi(x) }{a^2}
\\
& = & \sum_i \nabla_i^2 \phi + 0 a^2 \sum_i \nabla_i^4 \phi
- \frac{a^4}{72} \sum_i \nabla_i^6 \phi + \ldots \, .
\end{eqnarray}
which is free of $O(a^2)$ contamination.  We can get this form if
we replace the lattice Lagrangian term $-(\phi(x) - \phi(x-a \hat(i)))^2/2$
with
\begin{equation}
\frac{-1}{2} \left( \phi(x) - \phi(x-a\hat{i}) \right) \left(
 \frac{-1}{12} \phi(x + a\hat{i})
+ \frac{5}{4} \phi(x) - \frac{5}{4} \phi(x - a\hat{i}) + 
\frac{1}{12} \phi(x - 2a \hat{i}) \right) \, ,
\end{equation}
which means we can still expect to have a conserved energy, which must
however be defined in terms of the above gradient term.

This system is more complicated to evolve, perhaps doubling the number
of computations in an update.  What we have gained, though, is that the
lattice size necessary to model the desired system with a given accuracy
is greatly reduced; the number of lattice points needed is order the
square root of the number previously required.  This greatly reduces the
numerical demands of the evolution.

The implementation of an improved lattice action is not quite as simple
in nonabelian gauge theory, but the techniques involved have been worked
out in the Euclidean case some time ago\cite{Euclideanimprovement}.
The purpose of this 
paper is to develop and implement the correct improved Minkowski Hamiltonian,
and to use it to study the cutoff dependence of the system.

The paper is organized as follows.
Sec. \ref{KogutHam} will review the Kogut-Susskind 
implementation of Minkowski, classical lattice Yang-Mills theory, 
the lattice definition of
$dN_{CS}/dt$, the addition of a chemical potential for $N_{CS}$, and
the thermalization of the lattice system.  Everything in the section has
appeared in previous literature; I include it for didactic reasons, to
review necessary tools and ideas for the remainder of the paper.  In
Sec. \ref{improvedHam} I will construct a lattice Hamiltonian in continuum
time which is free of dimension 6 lattice artefacts and find a discrete
time algorithm for its evolution.  The kinetic (electric
field) part of the Hamiltonian is not diagonal in the obvious basis
for electric fields, so the modifications to the definition of $dN_{CS}/dt$,
chemical potential, and the thermalization algorithm are nontrivial.
With the Hamiltonian, chemical potential, and thermalization 
fully implemented, I
analyze the motion of $N_{CS}$ under a chemical potential in
SU(2) gauge theory, presenting numerical results in Sec. \ref{Results}.
I also compute the maximal Lyapunov exponent for the system.
Finally, Sec. \ref{Conclusion} concludes.

\section{Kogut-Susskind Hamiltonian and $N_{CS}$}
\label{KogutHam}

In this section I will review the construction of a lattice, Minkowski
model for classical Yang-Mills theory, developed by Kogut and Susskind 
\cite{KogutSusskind}.  I present a leapfrog algorithm for its update
and a definition for $\Delta N_{CS}/ \Delta t$ due to Ambjorn et. al.
\cite{Ambjornetal,AmbKras}, and a way to 
apply a chemical potential for $N_{CS}$
developed in \cite{paper1}.  Krasnitz has developed a 
thermalization algorithm for the system, based on a set of Langevin 
equations \cite{Krasnitz}; but I will present a much simpler thermalization
algorithm developed in \cite{paper1}, because it is this algorithm
I will extend to the improved model in Sec. \ref{improvedHam}.

\subsection{Degrees of Freedom and Hamiltonian}

To implement Yang-Mills theory on a lattice, Kogut and Susskind followed
Wilson's ideas \cite{Wilson}, taking very literally the nature of the
guage field as an affine connection for SU(2) objects in space.  If there
were a scalar field taking values in the fundamental representation of SU(2) 
at each point on a lattice,
the gauge field should provide a rule for parallel transporting the
field from lattice site to lattice site.  In the continuum the rule is
that the value of $\phi(x + a \hat{i})$, parallel transported to the 
point $x$, is
\begin{equation}
\phi_x (x+a \hat{i}) = {\cal P} \exp \left( \int_{x+ a \hat{i}}^{x} 
\frac{i \tau^a}{2} A_a^{j}(y) dy_j \right) \phi(x + a \hat{i}) \, ,
\end{equation}
with the path ordering placing points near $x$ to the left.  (Here and
throughout I will use nonrelativistic notation and a positive metric on
the space indicies.)
The lattice analog is to have a matrix $U \in$SU(2) which lives on the 
link between the lattice sites, so the parallel transported object is
just $U_i(x) \phi(x + a \hat{i})$.  The left index of $U$ lies at the 
basepoint of the link 
and the right index lies at its endpoint, so to perform a
gauge transform in which $\phi(x)$ is rotated by $G \in $SU(2) to $G\phi$,
the link $U_i(x)$ is rotated to $G U_i(x)$ and the link
$U_i(x - a \hat{i})$ becomes $U_i(x -a \hat{i}) G^{\dagger}$; 
so transporting $G \phi(x)$
back to $x-a \hat{i}$ gives $U_i(x - a \hat{i}) G^{\dagger} G \phi(x)$,
which is unchanged, as it should be, by the guage transform at $x$.
To forward transport $\phi(x)$ to $x + a\hat{i}$ we multiply by 
$U_i^{\dagger}(x)$.  Note that my convention is to label the link
by its basepoint, so $U_i(x)$ runs from $x$ to $x + a \hat{i}$; also from
now on I will drop $a$, measuring everything in lattice units, 
exept in those cases where it is useful to include it to count dimensions.
I will also drop hats where the meaning is obvious, ie $x+i$ means
$x + a \hat{i}$.

For the time being I will consider a spatial lattice but with continuous
time.  The connection in the time direction then remains a Lie algebra
element $A_0^a$ living at each lattice point, with parallel transport
of a field $\phi(x,t)$ to time $t + \Delta t$ given by
\begin{equation}
\phi_{t+ \Delta t}(x,t) = {\cal P} \exp \left( \int_t^{t + \Delta t} 
i \tau^a A_0^a \right) \phi(x,t)
\end{equation}
where the path ordering places earlier times to the right.  $A_0$ transforms
under a gauge change as $A_0 \rightarrow G(x,t) A_0(x,t) G^{\dagger}(x,t)
+ \partial_0 G(x,t)$.

The field strength is a curvature tensor and it reflects the failure of
an object, on transport around a closed curve, to return to its original
value.  The smallest nontrivial closed curve is the square plaquette, so
$F_{ij} \neq 0$ in the continuum case corresponds to the product of the
links around a plaquette,
\begin{equation}
 U_j (x) U_i (x + j) U^{\dagger}_j ( x + i) U^{\dagger}_i(x) 
\equiv U_{\Box}(x)
\end{equation}
not being the identity.  A reasonable definition of the magnetic field
strength is the Lie algebra element
\begin{equation}
\frac{1}{2} {\rm Tr} -i \tau^a U_{\Box}(x) \, ,
\end{equation}
which, although it transforms under gauge change as an adjoint object at
the point $x$, is really associated with the plaquette lying in the
positive $i$ and $j$ directions from $x$.  The continuum Lagrangian depends
on the $F_{ij}$ as $\int F_{ij} F_{ij}/4$, which is reproduced at leading 
order of a weak field expansion by 
\begin{equation}
-L_{U} = \sum_{\Box} 1 - \frac{1}{2} {\rm Tr} U_{\Box}
\end{equation}
where the sum is over all plaquettes in the lattice.
(The coefficient does not agree with the continuum one, but I will account
for this when I thermalize the system.)

Next consider the electric field part of the Lagrangian.  It is convenient
to write the time derivative of a link variable as
\begin{equation}
\frac{dU_i(x)}{dt} = i \tau^a \dot{U}_i^a(x) U_i(x) 
\end{equation}
and to expand the Lie algebra element $\dot{U}_i^a(x)$ as
\begin{equation}
\dot{U}_i^a(x) = A_0(x) - U_i(x)A_0(x+i)U^{\dagger}_i(x) + E_i^a(x) \, .
\end{equation}
The field $E_i^a$ transforms under gauge transformations as an adjoint
object and can be recognized in the weak field limit as the electric
field; alternately we can notice that it is the small $\Delta t$ limit
of a plaquette in the $i$ and time direction, traced against the Lie 
algebra and divided by $\Delta t$.  
The electric part of the Lagrangian should be
\begin{equation}
L_E = \sum_{x,i} \frac{ E_i^a(x) E_i^a(x)}{2} \, .
\end{equation}

Now we can derive Lagrange equations of motion.  Note first that the
time derivative of $A_0$ never appears, so $A_0$ enters the Lagrangian
like a Lagrange multiplier.  Varying with respect to $A_0$ gives 
\begin{equation}
\frac{\partial L}{\partial A_0^a(x)} = 0 = \sum_j \left( E_j^a(x) - 
E_j^b(x-i)   \frac{1}{2} {\rm Tr} -i \tau^a 
U^{\dagger}_j ( x - j) i \tau^b U_j(x-j) \right) \, ,
\label{Gaussconstraint}
\end{equation}
where the second term is just the $E$ field on the link leading into $x$,
parallel transported so its indicies reside at the point $x$.  This 
equation is a constraint, called the Gauss constraint because it 
corresponds to the continuum condition that $D_j E_j = 0$, which is
Gauss' Law.  If we initialize the evolution by choosing values for
the $U$ and $E$, then the Gauss constraint is a restriction on the 
allowed initial conditions.

The equations of motion derived by varying with respect to a link are
\begin{equation}
\frac{dE_i^a(x)}{dt} = - DF_i^a(x) \, , \quad
DF_i^a(x) \equiv \sum_{4 \Box_i(x)} \frac{1}{2} {\rm Tr} -i \tau^a 
U_{\Box} \, ,
\label{YMeqmo}
\end{equation}
where $4 \Box_i(x)$ refers to the 4 plaquettes which contain the link leaving
$x$ in the $i$ direction, with orientation chosen so that $U_i(x)$ and
not $U^{\dagger}_i(x)$ appears, and beginning and ending at the point $x$.
I use the notation $DF_i^a(x)$ because it is compact and remeniscient of
the equivalent continuum quantity, $(D_j F_{ji})^a (x)$.
Note that Eq. (\ref{YMeqmo}) automatically preserves the Gauss
constraint if the initial conditions satisfy it, because the time derivative
of Eq. (\ref{Gaussconstraint}) will, according to Eq. (\ref{YMeqmo}), 
consist of the sum of traces over plaquettes, and every plaquette which
contributes to one $E$ contributes to a neighbor with opposite sign.
This exact satisfaction is necessary because the evolution would not
make any sense if it excited the modes constrained to be zero; in effect
we keep track of more $E$ fields than are actually dynamical
degrees of freedom, and we must make sure that only the dynamical degrees
of freedom actually get excited.

Note that Eq. (\ref{YMeqmo}) does not actually uniquely specify the 
evolution of the system.  $E_i^a(x)$ does not give us $\dot{U}_i^a(x)$,
and the equations say nothing about how to evolve the value of $A_0$.
This is because the evolution should not in fact be uniquely specified,
yet; we have the freedom to make arbitrary time dependent gauge 
transformations.  To write a numerical update algorithm we have to decide
how to handle this freedom; the simplest and easiest choice is to pick the
gauge such that $A_0^a$ is everywhere zero.  The system can then
be considered as a constrained Hamiltonian system with coordinates $U$ and
conjugate momenta $E$.  The Hamiltonian is just $L_E - L_U \equiv H_E + H_U$.

\subsection{Numerical evolution of system}

One way to evolve this system is to use a leapfrog algorithm in which
the $U$ fields are defined at integer multiples of $\Delta t$, and
the $E$ are taken to be Lie algebra elements living at half time steps;
the rule is 
\begin{eqnarray}
E_i^a(x,t + \Delta t /2) & = & E_i^a(x,t - \Delta t /2 ) 
- \Delta t DF^a_i(x,t) \, \\ 
 U_i(x,t + \Delta t) & = & \exp( i \tau^a E^a_i(x) \Delta t) U_i(x) \, .
\end{eqnarray} 
This choice still identically preserves the Gauss constraint \cite{Krasnitz},
whether or not the parallel transport involved in the definition of the
Gauss constraint is performed with the past or future $U$.  The reason
is that $E$ commutes with its own exponent, so the parallel transport is
the same for the updated $U$ as for the $U$ before update.  This algorithm
forms the basis of the work in \cite{Krasnitz,AmbKras,paper1}.  

An alternate evolution algorithm developed in \cite{Ambjornetal}, 
which is closer to what I will need for
the improved Hamiltonian system, is to begin with a discrete time lattice,
so the $A_0$ fields are also connection matricies and the $E$ fields are
plaquette matricies for plaquettes with one space and one time step.  The
action is
\begin{equation}
I = \left( \sum_{\Box_s} - \frac{1}{(\Delta t)^2} \sum_{\Box_t} \right)
\left[  1 - \frac{1}{2}
{\rm Tr} U_{\Box} \right]  \, ,
\end{equation}
where $\Box_s$ are plaquettes entirely in space and $\Box_t$ are plaquettes
in one space and one time direction.
The equations of motion in the temporal gauge follow immediately; 
\begin{eqnarray}
U_i(x,t + \Delta t) & = & E_i(x,t + \Delta t /2) U_i(x,t) \, ,
\\
\frac{1}{2}{\rm Tr} -i \tau^a E_i(x,t + \Delta t /2) 
& = & \frac{1}{2}{\rm Tr} -i \tau^a E_i(x,t - \Delta t /2) 
- (\Delta t)^2 DF_i^a(x) \, .
\end{eqnarray} 
The only differences between the two update algorithms are that 
matrix exponentiation is
replaced by inverting $(1/2) {\rm Tr} -i \tau^a$, and
the definition of energy is changed from $E^a E^a/2$ to
$(1 - \sqrt{(\Delta t)^2 E^a E^a})/(\Delta t)^2$.  These changes are
quadratic in $|E^a| \Delta t$, which is no larger than the
natural level of error of the leapfrog algorithm, so the difference is
unimportant.

For finite $\Delta t$ there is no identically preserved quantity which
can be understood as a system energy; but a close reasonable approximation
is the quantity
\begin{equation}
{\rm Energy}(t) = \sum_{\Box} 1 - \frac{1}{2} {\rm Tr} U_{\Box}(t)
	+ \sum_{x,i} \frac{ (E^a_i(x,t + \Delta/2))^2
	+  (E^a_i(x,t - \Delta/2))^2 }{4}
\end{equation}
for the first update algorithm and
\begin{eqnarray}
{\rm Energy}(t) & = & \sum_{\Box} 1 - \frac{1}{2} {\rm Tr} U_{\Box}(t)
\nonumber \\
	& + & \! \frac{1}{2 (\Delta t)^2 } \sum_{x,i}  \left( 
	1 - \frac{1}{2} {\rm Tr} E_i(x , t + \Delta t/2) 
	+ 1 - \frac{1}{2} {\rm Tr} E_i(x , t - \Delta t/2) \right)
\end{eqnarray}
for the second.  
In each algorithm, the energy is preserved in the mean, with fluctuations
suppressed by $(\Delta t)^2$ and further washed out because they generically
add incoherently between different degrees of freedom.  The 
central value is absolutely stable, because the update algorithm has two 
important properties: (1) it bases the evolution of the system on the
minimum necessary amount of state information; and (2) it is exactly 
time reversal invariant (to evolve the system backwards, flip the sign
of $E$ in the first version or take Hermitian conjugate in the second).
It is therefore impossible that the energy should tend to rise exponentially,
since under time reversal it would tend to shrink exponentially, but the
algorithm is the same.  further, by good fortune the roundoff errors
which are inevitable in a numerical implementation add incoherently and
do not tend to increase the energy.  The stability of the system energy
is necessary for long Hamiltonian evolutions and will also allow checks
that the chemical potential method is behaving properly.

\subsection{Magnetic field and $N_{CS}$}

\begin{figure}
\centerline{\psfig{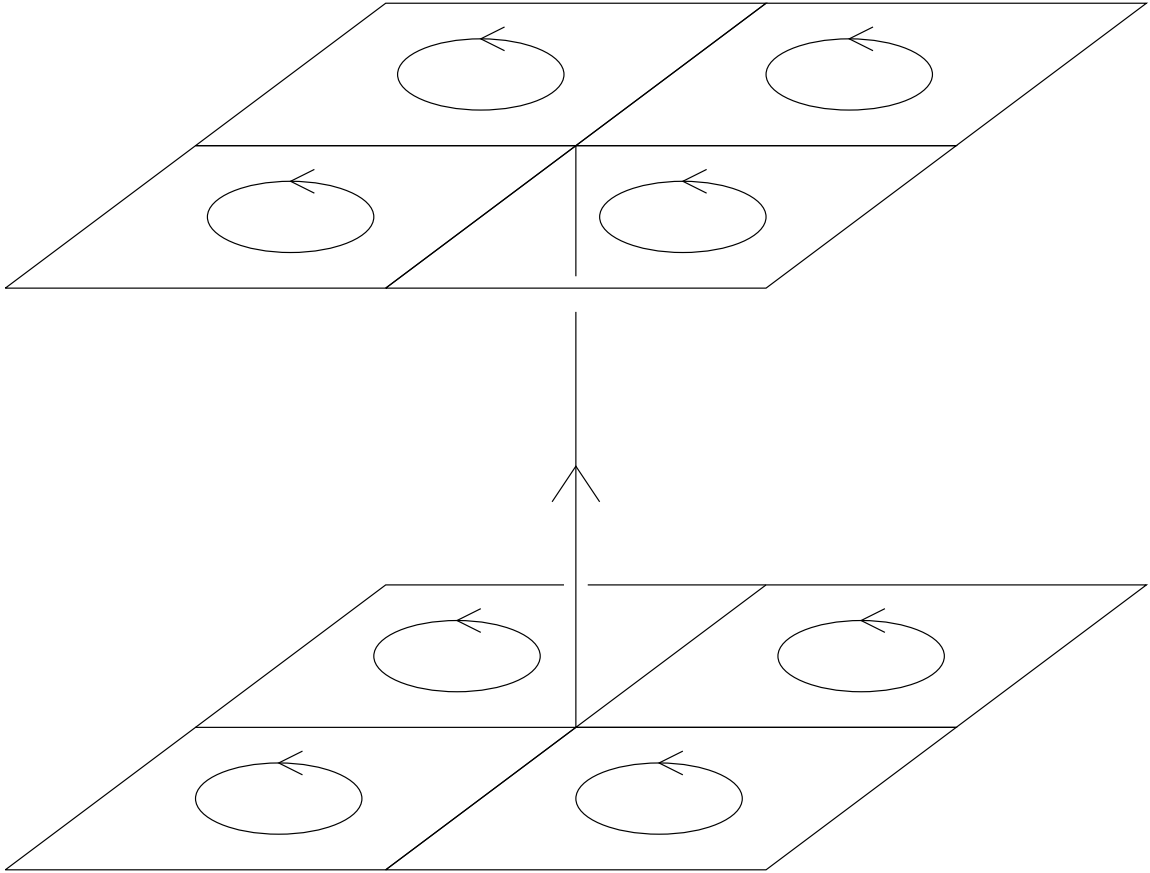}}
\caption{The 8 plaquettes contribute to the magnetic field along the
vertical link in the unimproved Hamiltonian system.
\label{8plaquette}
}
\end{figure}

To track the evolution of $N_{CS}$ in the realtime evolution of the Yang-Mills
fields described in the last section it is necessary to develop a 
definition of $dN_{CS}/dt$.  (It is impossible to define $N_{CS}$ directly
on a lattice, see \cite{paper1}).  In the continuum the relation is
\begin{equation}
\frac{dN_{CS}}{dt} = \int d^3x \frac{E^a_i(x) B^a_i(x)}{8 \pi^2} \, .
\end{equation}
We have already found lattice analogs of $E_i$ and $F_{jk} \sim B_i$.  
$E_i$ is associated with a link, and $F_{jk}$ 
is associated with a plaquette in the $j,k$ directions
(cyclic permutation implied), and the simplest definition of $d N_{CS}/dt$
which is invariant under the cubic point group is
\begin{equation}
\frac{dN_{CS}}{dt} = \sum_{x,i} \frac{E_i^a(x) B_i^a(x)}{2 \pi^2} \, ,
\qquad
B_i^a(x) \equiv \frac{1}{8} \sum_{8 \Box}
\frac{1}{2} {\rm Tr} -i \tau^a U_{\Box} 
\end{equation}
where $8 \Box$ means a sum over the 8 plaquettes which are orthogonal to
the link $x,i$ and touch one of its endpoints.  The plaquettes should
be traversed according to the righthand rule, starting and ending at the
point along the link, see Figure \ref{8plaquette}; and parallel transport
of the plaquettes which touch the endpoint of the link back to the 
basepoint is implied.  In the discrete time, leapfrog algorithm evolution of
the system the appropriate choice for $\Delta N_{CS}$ from one update is
\begin{equation}
2 \pi^2 \Delta N_{CS} = \Delta t \sum_{x,i} \frac{E^a_i(x,t + \Delta t/2)
+ E^a_i(x, t - \Delta t/2)}{2} B^a_i(x,t) \, ,
\end{equation}
where $E_i^a(x)$ should be replaced by $(1/2 \Delta t) {\rm Tr} -i \tau^a
E_i(x)$ for the algorithm used by Ambjorn et. al.

The coefficient is $2 \pi^2$ rather than $8 \pi^2$, as it is in the 
continuum, because the continuum equations are based on fields defined
in terms of $\tau^a/2$, and throughout I use $\tau^a$ in the discrete
system.  Similarly the $g^2$, which can appear in the continuum equations
depending on the normalization of $A_{\mu}$, does not appear here,
but will appear in the definition of lattice temperature instead.

The definition of $\Delta N_{CS}$ above corresponds to that used in
\cite{AmbKras,paper1}.  The definition used in \cite{Ambjornetal} 
was simpler, but not invariant under the cubic point group; so although
it does correspond to $N_{CS}$ in the continuum limit it is contaminated
by nonrenormalizeable operators already at dimension 5, whereas the
symmetric definition is only contaminated at dimension 6.

\subsection{Chemical potential for $N_{CS}$}
\label{chempot}

A particularly efficient way to study the rate of topology change in
the symmetric electroweak phase is to apply a chemical potential $\mu$ for
$N_{CS}$ and to measure $\dot{N}_{CS} T /\mu$.  To do so it is necessary
to develop a definition for the chemical potential, since there is
no good lattice analog to the continuum definition of $N_{CS}$.  The 
most naive solution is that a chemical potential modifies the dynamics
to
\begin{eqnarray}
\frac{dU_i(x)}{dt} & = & i E^a_i(x) \tau^a U_i(x)
\\
\frac{dE_i^a(x)}{dt} & = & -DF_i^a(x) - \frac{ \mu B_i^a(x)}{2 \pi^2}
\end{eqnarray}
which look just like the corresponding 
continuum equations.  However, this does not
work, because the $B$ term in the second equation does not preserve 
the Gauss constraint, ie
\begin{equation}
\sum_i \left( B_i^a(x) - B_i^a(x-i) \right) \neq 0
\label{Bianchifails}
\end{equation}
(where $B_i^a(x-i)$ should be parallel transported as an adjoint object along
the link leading to $x$).  The corresponding 
continuum expression, $D_i B_i$, is zero, by the Bianchi identity; but
the proof of the Bianchi identity depends on continuity of the $A$ fields,
and the notion of continuity is lost on the lattice.  
Eq. (\ref{Bianchifails}) is zero in abelian gauge theory.  For SU(2) in 
the low temperature or small $a$ limit,
the violations are higher order in temperature
than $B$, but nevertheless the failure of Eq. (\ref{Bianchifails}) to
equal zero means that the proposed implementation will spoil the Gauss
constraints, exciting unphysical modes.  As I discuss in \cite{paper1},
this makes the results unreliable.

The solution is to take very literally the constraints.  What they mean
is that we are not really living in the phase space of all $E$, $U$, but
on the constraint subspace, where the $E$ satisfy Gauss' law.  The only
actual dynamical fields are the linear combinations of $E$ which are
orthogonal to the Gauss constraints; if we could, we would express the 
system and its evolution in terms only of these.  

Consider the most obvious basis of electric fields, where a basis element
is a specification of a lattice point $x$, a direction $i$, and a Lie
algebra element $a$.  I will write this basis as $\{ E_{\alpha} \}$.
This basis has the
important property that, in terms of the inner product defined by 
$2 H_E$, it is orthonormal.  I write the Gauss constraints in this 
basis as $G_{\alpha} =  c_{\alpha \beta} E_{\beta} = 0$, 
where note that the first index
of $c$ only runs over the $3N^3$ Gauss constraints ($N$ the number of
lattice points on a side of a cubic lattice), while the second runs over
the $9N^3$ independent electric field 
basis elements.  (Sums on repeated greek indicies are always
implied.)  Note also that some of the
coefficients $c_{\alpha \beta}$ are $U$ dependent, because the $E$ fields
on the links leading into a point must be parallel transported to the 
point.  Now assuming that all Gauss constraints are 
independent\footnote{in the abelian theory with periodic boundary conditions
this is in fact not the case; there is precisely one linear dependence
between the Gauss constraints.  But for generic $U$ the Gauss constraints
in the nonabelian theory are independent.} we can find a different
orthonormal basis in which $c_{\alpha \beta}$ is nonzero only for
$\beta \leq 3N^3$, so the basis can be partitioned into two subsets, the
$E^c$, which are an orthonormal basis for the Gauss constraints, and
the $E^*$, which are the dynamical degrees of freedom.  
Now the definition
of the discrete $dN_{CS}/dt$ can be written as
\begin{equation}
2 \pi^2 \frac{dN_{CS}}{dt} = E^c \cdot B + E^* \cdot B \, .
\end{equation}
Since we are only interested in the situation in which all $E^c$ are
exactly zero, we may as well define $dN_{CS}/dt$ as
\begin{equation}
2 \pi^2 \frac{dN_{CS}}{dt} = E^* \cdot B \, ,
\end{equation}
and hence the action of the chemical potential should be to modify the
$E^* $ in proportion to the parallel component of $B$, but to leave the
$E^c$ alone.  In other words, if we write the orthogonal transformation
between the $E^c,E^*$ basis and the $E_{\alpha}$ basis as
\begin{equation}
E^*_{\beta} = a_{\beta \alpha}E_{\alpha} \, ,  \quad 
E^c_{\beta} = b_{\beta \alpha}E_{\alpha} \,
\end{equation}
where orthogonality ensures
\begin{equation}
 a_{\alpha \beta} a_{\gamma \beta} = \delta_{\alpha \gamma} \, ,
 \quad 
 b_{\alpha \beta} b_{\gamma \beta} = \delta_{\alpha \gamma} \, ,
\quad
 a_{\beta \alpha} a_{\beta \gamma} 
+  b_{\delta \alpha} b_{\delta \gamma} = \delta_{\alpha \gamma}
\, ,
\end{equation}
then the update rule should be
\begin{equation}
\frac{d E_{\alpha}}{dt} = -DF_{\alpha} - \frac{\mu}{2 \pi^2} 
 a_{\beta \alpha} a_{\beta \gamma} B_{\gamma}
\label{rightway}
\end{equation}
or equivalently
\begin{equation}
\frac{d E_{\alpha}}{dt} = -DF_{\alpha} - \frac{\mu}{2 \pi^2} 
\left( B_{\alpha} -  b_{\beta \alpha} b_{\beta \gamma}
B_{\gamma} \right) \, .
\label{rightwaytoo}
\end{equation}
This rule identically preserves the Gauss constraint while reflecting
the meaning we want for a chemical potential for $N_{CS}$, that it changes
electric fields in proportion to the parallel magnetic field component.
Particularly important, it will satisfy the property that the change
in the system energy 
\begin{equation}
\delta H = E \cdot \delta E = E^* \cdot \delta E^* 
= - \frac{\mu E^* \cdot B}{2\pi^2}
\end{equation}
is $- \mu N_{CS}$, as it should be.  This allows a check of the algorithm
by comparing the accumulated energy to $N_{CS}$.  (Note that this is where
it is essential that the algorithm without chemical potential should
preserve system energy, in the mean, to high precision.)

The problem with this update rule lies in finding the orthogonal 
transformation to the basis which splits into constraints and
dynamical fields, which requires diagonalizing the matrix 
$c_{\alpha \beta}$.  Even if the diagonalization procedure could be
done efficiently, using it via Eq. (\ref{rightway}) or (\ref{rightwaytoo})
would take $O(N^6)$ computations, which is prohibitive.  However,
notice that Eq. (\ref{rightwaytoo}) shows that the correct algorithm
is to perform the naive implementation, but then to orthogonally project
back to the constraint surface.  This can be done approximately, but
very accurately, by an algorithm with $O(N^3)$ computations
per update.  

The idea is the following \cite{paper1}; the most general way of moving
orthogonally to the constraint surface is to change $E_{\alpha}$ by
\begin{equation}
\delta E_{\alpha} = \kappa_{\beta} c_{\beta \alpha} \, ,
\end{equation}
with $\kappa$ completely arbitrary.  (It is essential that the motion be
strictly orthogonal to the constraint surface so the dynamical fields are
not changed.)  The right choice of $\kappa$ is
the one which will change $E$ so as to satisfy all Gauss constraints.
A reasonable first guess, which is guaranteed to reduce the sum of
squares of Gauss constraint violations, is 
$\kappa_{\alpha} = - \gamma c_{\alpha \beta} E_{\beta}$, $0 < \gamma < 1/6$,
which is a step in a dissipative or quenching algorithm for the Gauss
constraints.  (The condition $\gamma < 1/6$ is demanded by stability.)
A crude approximate 
projection algorithm is to do one application of this quench at
each time step.  A more refined algorithm is to guess that the 
correction should be almost the same as in the previous timestep, and
to first apply $\kappa_{\beta} = m \kappa_{\beta, {\rm previous \; step}}$,
where $m<1$ ensures algorithm stability and 
$ \kappa_{\beta, {\rm previous \; step}}$ is the sum of all corrections 
applied at the previous timestep.  Then the dissipative step is applied
twice, with different stepsizes $\gamma_1$ and $\gamma_2$. I find that 
using the choices
$m = 1 - 1.6 \Delta t$, $\gamma_1
= 5/24$, $\gamma_2 = 5/48$ is stable and highly efficient. 
(Note that when two applications of the quench algorithm are performed in
a row, the stability criterion changes, see \cite{paper1}.)
Applying this approximate projection algorithm requires
about as many floating point operations as determining the $DF_{\alpha}$,
so it is not a great strain on algorithm efficiency.
It is also close enough to an exact projection that the residual failure
has virtually no effect on the results \cite{paper1}.

\subsection{Thermalization}
\label{unimp_thermalization}

Finally I discuss the thermalization of the lattice Yang-Mills system
developed in this section.
What a thermalization algorithm needs to do is to draw a set of $E$, $U$
from the constraint submanifold of the phase space of $E$, $U$, choosing
with weight $\exp - \beta_L H$ times the natural measure on the constraint
surface.  This definition of $\beta_L$ corresponds to 
$\beta_L = 4 \beta_{\rm continuum} / (g^2 a)$, with $a$ the lattice
spacing in the same physical units as $\beta_{\rm continuum}$.
(The factor of $g^2$ is the usual Yang-Mills wave function normalization and
the factor of 4 arises because I work in terms of $\tau^a$ rather
than $\tau^a/2$, which are used in the definition of the continuum fields.)
It is possible to perform the thermalization 
for a general constraint manifold,
using a set of Langevin equations \cite{Krasnitz}; but by using the
special properties that the subspace of $E$ fields at a fixed value of $U$
is a vector space, that the constraints are linear, and that the 
Hamiltonain is separable into a part which depends only on $U$ and a
part which is quadratic in $E$ at fixed $U$, it is possible to find a
simple and efficient thermalization algorithm \cite{paper1}.

The idea is as follows; while the $U$ dependence of the Hamiltonian,
$H_U$, is quite nontrivial, all of its complexities appear in the
equations of motion; and the equations of motion preserve the thermal
probability distribution, while mixing excitation between the $E$
and $U$ fields.  Since we must implement the equations of motion anyway,
it is best to use them to thermalize the system.  All we need is a
thermalization algorithm for the $E$ fields at fixed $U$; we then repeatedly
draw the $E$ fields from the fixed $U$ thermal distribution, evolve
the system to mix the thermalization between $E$ and $U$ fields, and
throw away the $E$ fields and repeat.  The idea is identical in spirit
to the ``molecular dynamics'' Monte-Carlo algorithm of Euclidean lattice
gauge theory.  The only complication here is that the $E$ fields must
satisfy constraints.

The thermalization of the $E$ fields is 
possible because the $E$ fields at fixed $U$ form a vector space,
the constraints are linear, and $H_E$ is quadratic in $E$.  In
terms of the orthogonal basis which partitions into $E^* $ and $E^c $,
the Hamiltonian is just $H_E =  E^*_\alpha E^*_\alpha /2 +
 E^c_{\alpha} E^c_\alpha /2$; 
we just need to choose each $E^*_\alpha $ from the Gaussian distribution
\begin{equation}
{\cal P}(E^{*}_{\alpha}) = 
\sqrt{\beta_l/2\pi} \exp - \beta_L( E^*_\alpha )^2/2
\end{equation}
and set each $E^c$ to zero.  The easiest way to do this, given the difficulty
of finding the basis which partitions into $E^* $ and $E^c $, is to
set both equal to the Gaussian distribution, and then orthogonally
project to the constraint surface.  
To get the initial Gaussian distribution, I simply draw the
$E_{\alpha}$ from the same Gaussian distribution; because the desired basis
is gotten by orthogonal transformation from this natural basis, the resulting
distribution will also be Gaussian in the basis of $E^* $ and $E^c $.
The orthogonal projection is performed
by using the quenching step of the algorithm of the last section 
repeatedly until the projection is complete.  I find that order 100
applications reduce the Gauss constraint violation to the level of 
roundoff error (for single precision numerics).

In practice I find this algorithm to be very efficient.  For instance, the
total energy of the system approaches the thermal average value by
a factor of $1/2$ on each $E$ randomization, if the Hamiltonian evolutions
are taken to be of length $\simeq \beta_L$; within 7 or 8 applications
the energy has reached the level of random fluctuation expected in
a system of finite size, and the values of Wilson loops equal their
thermal averages to within expected thermal fluctuation.  
Two or three applications are enough to re-randomize a thermalized 
distribution.

This supplies all necessary equipment for the numerical study of classical
3-D Yang-Mills theory under the minimal action.

\section{Improved Hamiltonian}
\label{improvedHam}

In this section I construct an improved Hamiltonian for 
the classical lattice Yang-Mills theory.  I also present an improved 
definition of $dN_{CS}/dt$ and discuss how to implement a chemical
potential for this quantity, and how to thermalize the system.

\subsection{a false start}

The simplest guess for an improved Hamiltonian takes inspiration from 
the example of 
scalar theory presented in the introduction and replaces the gradient
term, $H_U$, with an improved term,
\begin{equation}
H_{U,{\rm improved}} = \frac{5}{3} \left( \sum_{\Box} 1 - \frac{1}{2}
{\rm Tr} U_{\Box} \right) - \frac{1}{12} \left( 
\sum_{\sqsubset \! \sqsupset} 1 - 
\frac{1}{2} {\rm Tr} U_{\sqsubset \! \sqsupset} \right) \, ,
\end{equation}
which is known from the study of Euclidean lattice gauge theory 
\cite{Euclideanimprovement}.  
Here $\sum_ {\sqsubset \! \sqsupset}$ refers to a sum
over all $1 \times 2$ rectangular plaquettes, of which there are $6 N^3$. 
The evolution of the system is the same as in the previous section,
except that the definition of $DF_i^a(x)$ changes to
\begin{equation}
DF_i^a(x) = \frac{5}{3} \left( \sum_{4 \Box} 
\frac{1}{2} {\rm Tr} -i \tau^a U_{\Box} \right) - \frac{1}{12} \left(
\sum_{12 \sqsubset \! \sqsupset } \frac{1}{2} {\rm Tr} -i \tau^a
U_{\sqsubset \! \sqsupset} \right)
\label{improvedDF}
\end{equation}
where $12 \sqsubset \! \sqsupset$ refers to the 12 rectangular plaquettes
containing the link $U_i(x)$, traversed so as to begin and end at $x$ and
to contain $U_i(x)$ and not $U^{\dagger}_i(x)$.

To see that this proposal fails, it is useful to consider a
sinusoidal excitation in a single Lie algebra element, and to compute
the dispersion relation.  Consider the infinite volume limit of the
lattice and start from the initial conditions 
\begin{equation}
U_i(x) = I \, , \qquad
E_i(x) = \epsilon \tau^a 
E_i \sin \left( k \cdot (x + \frac{1}{2} \hat{i}) \right)
\end{equation}
where $\epsilon$ is infinitesimal.  (The coordinate in the argument of sin
is the center 
of the link.)  The energy per site of this arrangement is $\sum_i (E_i)^2
\epsilon^2/2$.  An infinitesimal time $\delta t$ later some of this
energy has appeared as magnetic energy; the fraction is 
$\delta t^2 \omega^2$, and $\omega$ gives the dispersion relation.  At time
$a \delta t$, the $U$ fields are
\begin{equation}
U_i(x) \simeq I + i \tau^a E_i(x) \delta t
\end{equation}
and a lengthy but straightforward calculation gives the energy from square
plaquettes as
\begin{equation}
\frac{5}{3} \frac{\epsilon^2 \delta t^2}{2} \left( (E_1 s_2 - E_2 s_1)^2 +   
 (E_1 s_3 - E_3 s_1)^2 +   
 (E_3 s_2 - E_2 s_3)^2 \right) \, ,
\end{equation}
(where I introduce the shortand notation $s_i \equiv 2 \sin( a k_i /2 ) $),
the energy from rectangular plaquettes as
\begin{equation}
\frac{-1}{12}  \frac{\epsilon^2 \delta t^2}{2}
\left( (8 - s_1^2 - s_2^2)  (E_1 s_2 - E_2 s_1)^2  +
( 1 \leftrightarrow 3) + ( 2 \leftrightarrow 3)
 \right) \, ,
\end{equation}
and the Gauss constraint as
\begin{equation}
s_1 E_1 + s_2 E_2 + s_3 E_3 = 0 \, .
\end{equation}
by adding the square of the Gauss constraint, which is zero, to the 
energy in magnetic fields, I readily find that the dispersion relation is
\begin{eqnarray}
a^2 \omega^2  & = &  ( s_1^2 + s_2^2 + s_3^2 + \frac{1}{12} ( s_1^4 + s_2^4
+ s_3^4) ) 
\\
& + & \frac{ E_1^2}{E_1^2 + E_2^2 + E_3^2} \left( \frac{1}{12} s_1^2 ( s_2^2
+ s_3^2 - s_1^2) \right) +  ( 1 \leftrightarrow 2 ) + ( 1 \leftrightarrow 3 )
\\
& + & \frac{E_1 E_2}{E_1^2 + E_2^2 + E_3^2} \left( \frac{-1}{6}
s_1 s_2 ( s_1^2 + s_2^2 ) \right) + ( 1 \leftrightarrow 3 ) 
+ ( 2 \leftrightarrow 3 )
\, .
\end{eqnarray}

The contribution to $\omega^2$ from the first line is $k_i^2$ 
plus zero times terms of
form $a^2 k^4$, plus $O(a^4 k^6)$, and is 
therefore the correct dispersion relation, up to corrections which 
vanish as $a^4$ in comparison to the leading term.  
The other terms, which vanish when
$E_1 s_1 = E_2 s_2 = E_3 s_3 = 0$, spoil the dispersion relations by
introducing polarization dependent $O(a^2 k^4)$ contibutions, so the
evolution is not ``improved''.

An even simpler way of seeing that the Hamiltonian is not ``improved'' is
to consider the system's thermodynamics, which are determined by integration
of the fields over the weight $\exp - \beta_L H$.  Here it is convenient
to follow \cite{AmbKras} and introduce Lagrange multipliers $A_0^a(x)$ for
the Gauss constraints, so the Hamiltonian is
\begin{equation}
H = \sum_{i,x} \left( \frac{E^a_i(x) E^a_i(x)}{2} \right) 
+ \sum_{x} i A_0^a(x) \left( \sum_{i} E^a_i(x) - E^a_i(x-i) \right) + 
H_{U,{\rm improved}}
\end{equation}
where again parallel transport of $E^a_i(x-i)$ to the point $x$ is to
be performed along the link leading into $x$.  In terms of the notation
of Sec. \ref{chempot} I can write this as
\begin{equation}
H = \frac{1}{2} E_{\alpha} E_{\alpha} + i A_{0 \beta} c_{\beta \alpha} 
E_{\alpha} + H_{U, {\rm improved}} \, .
\end{equation}
If we are only interested in the values of gauge invariant operators
constructed from the link matricies then we can perform the integral over
the $E$ fields, which is now Gaussian, and gives
\begin{equation}
H = \frac{1}{2} A_{0 \alpha} c_{\alpha \beta} c_{\gamma \beta} A_{0 \gamma}
+ H_{U, {\rm improved}} \, .
\end{equation}

Note that $ A_{0 \beta } c_{\beta \alpha} E_{\alpha}$ is (performing the
sum on $\beta$ and expressing the sum on $\alpha$) 
$\sum_{x,i}E_i^a(x) (A_0^a(x) - A_0^a(x+i) )$, 
parallel transport of the forward $A_0$ implied, as usual.  So, the $A_0$
expression above can be untangled to read
\begin{equation}
H = \frac{1}{2} \sum_{x,i} (A_0^a(x) - A_0^a(x+i)) (A_0^a(x) - A_0^a(x+i))
+ H_{U, {\rm improved}} \, ,
\end{equation}
which is the lattice action of the finite temperature quantum theory
in the approximation of dimensional reduction,
with zero bare Debye mass.  It has the improved plaquette action, but
the unimproved scalar ($A_0$) gradient term, as a comparison with the
scalar theory presented in the introduction will show.  Hence the
Hamiltonian has not been ``improved,'' even at the level of thermodynamics,
and this failure has nothing to do with the choice for an improved
magnetic term.

\subsection{Hamiltonian and Gauss constraints}

The problem with the approach above is that, while I have removed 
nonrenormalizeable operators such as 
$a^2 \sum_{ij} (D_i F_{ij})^a (D_i F_{ij})^a$, I have left in the 
nonrenormalizeable operator $a^2 \sum_i (D_i E_i)^a (D_i E_i)^a $.
An easy way to see why is to recall how the minimal Hamiltonian system
could be interpreted as arising from an action on a lattice in both
space and time, consisting of all $1 \times 1$ plaquettes, in space and
time.  The $E^2$ term in $H$ is just the sum over the $1 \times 1$ plaquettes
with time links.  In the previous section, I included $1 \times 2$
plaquettes only if each direction were in space, but continued to use
a kinetic $E^2$ term derived only from the square plaquettes.  Instead I
should include rectangular plaquettes here too.  The result is that
the continuum time Hamiltonian should be
\begin{equation}
H_{E , {\rm improved}} = \frac{5}{3} \sum_{i,x} \frac{(E_i^a(x))^2}{2}
- \frac{1}{12} \sum_{i,x} \frac{ (2 E_i^a(x))^2 }{2}
- \frac{1}{12} \sum_{i,x} \frac{ ( E_i^a(x) + E_i^a(x + i))^2}{2} \, ,
\end{equation}
where the first term comes from square plaquettes, the second comes from
plaquettes which stretch 2 units in time, and the third comes from 
plaquettes which stretch 2 units in space, and I have already taken the
small time step limit.  As usual, $E_i^a(x+i)$ must be parallel transported
along the straight path.  The second term should be absorbed in the
first, changing its coefficient to $4/3$.  I have retained my earlier
definition of $E_i^a(x)$, so in the $A_0^a = 0$ gauge the update rule for
the $U$ is
\begin{equation}
\frac{dU_i(x)}{dt} = i \tau^a E_i^a(x) U_i(x) \, .
\end{equation}

The Hamiltonian is 
conveniently summarized by defining the operator $M$,
\begin{equation}
(ME)_i^a(x) = \frac{7}{6} E_i^a(x) - \frac{1}{12} 
\Big( E_i^a(x-i) + E_i^a(x+i) \Big)
\end{equation}
(parallel transports again implied), in terms of which the Hamiltonian is
\begin{equation}
\label{rightHamiltonian}
H = H_{U, {\rm improved}} 
+ \frac{1}{2} E_{\alpha} M_{\alpha \beta} E_{\beta} \, .
\end{equation}
The Gauss constraint is derived as before, by varying $A_0(x)$;
it is the sum of all plaquettes which traverse the time direction and
terminate at $x$, with appropriate signs:
\begin{eqnarray}
0 & = & \sum_i \left( \frac{4}{3} \Big( E_i^a(x) - E_i^a(x-i) \Big)
\right. \nonumber \\
&- & \left. \frac{1}{12} \Big( (E_i^a(x) + E_i^a(x+i)) 
- (E_i^a(x-i) + E_i^a(x-2i) )  \Big) \right)  \, ,
\end{eqnarray} 
or in somewhat more compact notation,
\begin{equation}
c_{\beta \alpha} M_{\alpha \gamma} E_{\gamma} = 0 \, .
\label{rightGaussconstraint}
\end{equation}

Now I will address the question of whether this Hamiltonian is 
improved in the sense of replicating the continuum evolution for smooth
fields without contamination from dimension 6 operators.
There are two general types of dimension 6, nonrenormalizeable operators
which can appear in cubic point group invariant 
lattice implementations of Yang-Mills
theory; there are those with two powers of the field strength tensor,
such as $\sum_{ij} (D_i F_{ij})^a (D_i F_{ij})^a$,
and terms which contain three powers of the field strength, like
$f_{abc} F_{ij}^a F_{jk}^b F_{ki}^c$.  In the
latter term, the indicies of the $F$ must differ, so that they represent
curvatures in different planes, since $f_{abc} F_{ij}^a F_{ij}^b$
automatically vanishes.  Since all the plaquettes used in 
Eq. (\ref{rightHamiltonian}) are planar, terms involving three powers of
the field strength will not appear at dimension 6.  I can therefore
restrict attention to the other sort of term, which appear already in
the Abelian theory and which modify the dispersion relations of 
weak disturbances which vary sinusoidally in space.
It is sufficient, then, to check whether the dispersion
relation for $\omega^2$ is free of $k^4 a^2$ terms in the small $k$ 
expansion.  It is
again sufficient to start a sinusoidal excitation with $U = I$ everywhere
and see how quickly the energy is transferred into the magnetic
term of the Hamiltonian.  For
\begin{equation}
E_i(x) = \epsilon \tau^a E_i \sin( k \cdot (x + i/2) )
\end{equation}
the electric energy is $\epsilon^2/2$ times
\begin{equation}
\frac{4}{3} \left( E_1^2 + E_2^2 + E_3^2 \right) 
- \frac{1}{12} \left( (4 - s_1^2) E_1^2 + ( 4 - s_2^2) E_2^2
+ ( 4 - s_3^2) E_3^2 \right) \, ,
\end{equation}
with the first term from the single $E$ term in the Hamiltonian and
the second term from the two $E$ term in the Hamiltonian.
Again I use the notation $s_i = 2 \sin k_i/2$.

The new Gauss constraint is
\begin{equation}
\left( 1 + \frac{1}{12} s_1^2 \right) E_1 s_1 + 
\left( 1 + \frac{1}{12} s_2^2 \right) E_2 s_2 + 
\left( 1 + \frac{1}{12} s_3^2 \right) E_3 s_3  = 0 \, ,
\end{equation}
and the magnetic energy at time $a \delta t$ is still
\begin{equation}
\frac{\epsilon^2 ( \delta t)^2}{2}
\left( 1 + \frac{1}{12} ( s_1^2 + s_2^2) \right) (E_1 s_2 - E_2 s_1)^2
+ (1 \leftrightarrow 3) + ( 2 \leftrightarrow 3 )
\end{equation}
as in the previous subsection.  Squaring the Gauss constraint, which
equals zero, and adding it to the magnetic energy term, I find that the
oscillation frequency is
\begin{eqnarray}
a^2 \omega^2 & = & \sum_i ( s_i^2 + \frac{1}{12} s_i^4 )
\\ & + & \frac{1}{144} \,
\frac{ s_1^2 s_2^2 (E_1 s_2 - E_2 s_1)^2 + ( 1 \leftrightarrow 3 )
+ ( 2 \leftrightarrow 3 ) }{ \left( 1 + \frac{s_1^2}{12} \right) E_1^2
 + \left( 1 + \frac{s_2^2}{12} \right) E_2^2
 + \left( 1 + \frac{s_3^2}{12} \right) E_3^2 }  \, .
\end{eqnarray}
The first term here is
\begin{equation}
\frac{8}{3} \sum_i (1 - \cos a k_i )  - \frac{1}{6} \sum_i (1 - \cos 2a k_i )
\end{equation}
which is the usual dispersion relation for an improved, scalar
Laplacian.  It contributes no $a^2 k^4$ terms to $\omega^2$, 
as can be readily verified by
expanding the cosines as power series.  The second term is polarization
dependent, but has 6 powers of $2 \sin a k/2$, so it only starts to
contribute at order $a^4k^6$, and the dispersion relation is therefore
free of the unwanted $a^2 k^4$ terms, and the Hamiltonian is free of
dimension 6 nonrenormalizeable operators.  Also note that the new Gauss
constraint demands that $E$ be transverse to a higher order in $k$ than the
unimproved Gauss constraint does.

I can also check the thermodynamics as before.  Introducing $A_0(x)$ as
a Lagrange multiplier for the constraint, the thermodynamics is 
governed by the probability distribution $\exp - \beta_L H$, with
\begin{equation}
H = H_{U , {\rm improved}} + \frac{1}{2} E_{\alpha} M_{\alpha \beta}
E_{\beta} + i A_{0 \beta} c_{\beta \gamma} M_{\gamma \alpha} E_{\alpha}
\end{equation}
which, on performing the Gaussian integration over $E$, becomes
\begin{equation}
H_{U,{\rm improved}} + \frac{1}{2} A_{0 \beta} c_{\beta \alpha}
M_{\alpha \gamma} c_{\delta \gamma} A_{0 \delta} \, .
\end{equation}
The $A_0$ term untangles to
\begin{equation}
 \frac{1}{2} (A_0(x) - A_0(x+i)) \left(
\frac{1}{12} A_0(x+2i) - \frac{5}{4} A_0(x+i) + \frac{5}{4} A_0(x)
- \frac{1}{12} A_0(x - i) \right) \, ,
\end{equation}
parallel transports again implicit.  The derivative term for $A_0$ 
is precisely the improved derivative term for the scalar theory discussed
in the introduction, made covariant.  Hence the Hamiltonian does give improved
thermodynamics.  This also points out a relation between the improvement
scheme for the plaquette action and for the discrete Laplacian.

\subsection{Numerical evolution of the improved system}

Rather than derive continuum time equations of motion from the Hamiltonian,
I will skip to finding a discrete time evolution algorithm.  Naively, 
I should base this on the action
\begin{eqnarray}
S & = & S_E - S_U
\nonumber \\
S_E & = & \frac{1}{(\Delta t)^2} \sum_{x,t} \left[ \frac{5}{3} \sum_{i}
1 - \frac{1}{2} {\rm Tr} U_{\Box_{i0}}
- \frac{1}{12} \sum_i 1 - \frac{1}{2} {\rm Tr} 
U_{\sqsubset \! \sqsupset_{i0}}
- \frac{1}{12} \sum_i 1 - \frac{1}{2} {\rm Tr} 
U_{\sqsubset \! \sqsupset_{0i}}
\right]
\nonumber \\
S_U & = & \sum_{x,t} \left[ \frac{5}{3} \sum_{i>j}
1 - \frac{1}{2} {\rm Tr} U_{\Box_{ij}}
- \frac{1}{12} \sum_{i>j} 1 - \frac{1}{2} {\rm Tr} 
U_{\sqsubset \! \sqsupset_{ij}}
- \frac{1}{12} \sum_{i>j} 1 - \frac{1}{2} {\rm Tr} 
U_{\sqsubset \! \sqsupset_{ji}}
\right] \, ,
\end{eqnarray}
which includes plaquettes which go two steps in time.  These plaquettes
prevent a nonrenormalizeable operator of form 
$(\Delta t)^2 (D_0 F_{0i})^a (D_0 F_{0i})^a$ and should make the evolution
fourth order in stepsize.  This is unnecessary if we take $\Delta t << 1$,
and is also undesirable since the inclusion of plaquettes which stretch two
steps in time makes the numerical evolution unstable.  The reason is that
the update rule derived by varying one link then determines $U$ based on
the 4 previous time slices, rather than just 2 as the minimal algorithm
did; since the Yang-Mills equations are second order, this
is too much state information, which opens the possibility (which is
realized) of an unstable growing mode.  Instead I choose $\Delta t$ small
enough to make those plaquettes unnecessary, and choose
\begin{equation}
S_E = \frac{1}{(\Delta t)^2} \sum_{x,t} \left[ \frac{4}{3} \sum_{i}
1 - \frac{1}{2} {\rm Tr} U_{\Box_{0i}}
- \frac{1}{12} \sum_i 1 - \frac{1}{2} {\rm Tr} 
U_{\sqsubset \! \sqsupset_{0i}}
\right] \, .
\end{equation}

As before, I will define $E_i(x , t + \Delta t/2)$ as the square 
plaquette, so in the gauge $U_0(x) = I$,
\begin{equation}
U_i(x,t + \Delta t) = E_i(x, t + \Delta t/2) U_i(x , t) \, .
\end{equation}
The Gauss law is that the sum of Lie algebra traces of all plaquettes 
containing a time link be zero, which in the same gauge is
\begin{eqnarray}
0 & = & \sum_i \left[ \frac{4}{3} \left( 
\frac{1}{2} {\rm Tr} -i \tau^a U_i(x,t+\Delta t)
U^{\dagger}_i(x,t) - 
\right. \right.  \\ & & \qquad  \quad \left.
\frac{1}{2} {\rm Tr} -i \tau^a U^{\dagger}_i(x-i,t)
U_i(x-i,t + \Delta t) \right) - 
\nonumber \\ &&
 \frac{1}{12} \left( 
\frac{1}{2} {\rm Tr} -i \tau^a U_i(x,t+\Delta t) U_i(x+i,t + \Delta t)
U^{\dagger}_i(x+i , t) U^{\dagger}_i(x , t)
- \right.
\nonumber \\ && \qquad \left. \left.
\frac{1}{2} {\rm Tr} -i \tau^a U^{\dagger}_i(x-i,t) 
U^{\dagger}_i(x-2i,t) U_i(x-2i , t + \Delta t) U_i ( x-i , t + \Delta t)
\right) \right] \nonumber
\end{eqnarray}
or
\begin{eqnarray}
0 & = & \sum_i \left[ \frac{4}{3} \left( 
\frac{1}{2} {\rm Tr}-i \tau^a E_i(x,t+\Delta t/2)
- \right. \right.
\nonumber \\ && \left. \qquad
\frac{1}{2} {\rm Tr} -i \tau^a U^{\dagger}_i(x-i,t) E_i(x-i,t + \Delta t/2)
U_i(x-i,t) \right)  -
\nonumber \\ &&
\frac{1}{12} \bigg( \frac{1}{2} 
{\rm Tr} -i \tau^a E_i(x,t+\Delta t/2) U_i(x,t)E_i(x+i,
t+\Delta t/2) U^{\dagger}_i(x,t)
-
\nonumber \\ && \qquad
\frac{1}{2} {\rm Tr} -i \tau^a U^{\dagger}_i(x-i,t) U^{\dagger}_i(x-2i,t) 
E_i(x-2i,t+\Delta t/2) \times
\nonumber \\ &&  
\qquad \qquad  \qquad U_i(x-2i,t) E_i(x-i,t + \Delta t/2) U_i(x-i,t) \bigg)
\bigg] \, .
\end{eqnarray}
Implicitly parallel transporting using the connection at time $t$,
this is
\begin{eqnarray}
0 & = & 
\sum_i \left[ \frac{4}{3} \left(
\frac{1}{2} {\rm Tr} -i \tau^a E_i(x,t + \Delta t/2)
- \frac{1}{2}{\rm Tr} -i \tau^a E_i(x-i,t + \Delta t/2) \right)
- \right. 
\nonumber \\ &&
\frac{1}{12} \left( \frac{1}{2} {\rm Tr} -i \tau^a E_i(x,t + \Delta t/2) 
E_i(x+i, t + \Delta t/2)
- \right.
\nonumber \\ & & \left. \left. \qquad
\frac{1}{2}
{\rm Tr} -i \tau^a E_i(x-2i,t + \Delta t/2) E_i(x-i,t + \Delta t/2 ) 
\right) \right] \, ,
\end{eqnarray}
whereas if the parallel transports are performed using the connection
at time $t + \Delta t$, this is
\begin{eqnarray}
0 & = & 
\sum_i \left[ \frac{4}{3} \left( 
\frac{1}{2} {\rm Tr} -i \tau^a E_i(x,t + \Delta t/2)
- \frac{1}{2} {\rm Tr} -i \tau^a E_i(x-i,t + \Delta t/2) \right)
- \right. 
\nonumber \\ &&
\frac{1}{12} \left( \frac{1}{2} {\rm Tr} -i \tau^a E_i(x+i,t + \Delta t/2) 
E_i(x, t + \Delta t/2)
- \right.
\nonumber \\ & & \left. \left. \qquad
\frac{1}{2} {\rm Tr} -i \tau^a 
E_i(x-i,t + \Delta t/2) E_i(x-2i,t + \Delta t/2 ) 
\right) \right] \, .
\end{eqnarray}

The equations of motion, derived by varying one spatial link, are also
somewhat more complicated.  In terms of the definition of $DF_i^a(x)$
appearing in Eq. (\ref{improvedDF}), and implicitly parallel
transporting objects using the connections at time $t$ and the straight
line paths, the update rule is
\begin{eqnarray}
&& \frac{4}{3} \left( \frac{1}{2} {\rm Tr} -i \tau^a E_i(x,t + \Delta t/2)
\right) - 
\nonumber \\ &&
\frac{1}{12} \left( 
\frac{1}{2} {\rm Tr} -i \tau^a E_i(x , t+ \Delta t/2) 
E_i(x + i , t + \Delta t/2) + 
\right. \nonumber \\ && \left. \qquad
\frac{1}{2} {\rm Tr} -i \tau^a E_i(x-i , t + \Delta t/2)
E_i(x , t + \Delta t/2) \right) =
\nonumber \\
&& \frac{4}{3} \left( \frac{1}{2} {\rm Tr} -i \tau^a E_i(x,t - \Delta t/2)
\right) -
\nonumber \\ &&
 \frac{1}{12} \left( 
\frac{1}{2} {\rm Tr} -i \tau^a E_i(x + i , t - \Delta t/2) 
E_i(x , t - \Delta t/2) +
\right. \nonumber \\ && \left. \qquad
\frac{1}{2} {\rm Tr} -i \tau^a E_i(x , t - \Delta t/2)
E_i(x-i , t - \Delta t/2) \right) - (\Delta t)^2 DF_i^a(x , t) \, .
\label{updaterule}
\end{eqnarray}

Writing this equation in the convenient shorthand
\begin{equation}
(ME)_i^a(x,t + \Delta t/2) = (ME)_i^a(x , t- \Delta t/2) 
- ( \Delta t)^2 DF_i^a(x) \, ,
\end{equation}
one can see that the Gauss constraint is
\begin{equation}
\sum_i ( (ME)_i^a(x,t + \Delta t/2) - (ME)_i^a(x-i , t + \Delta t/2) ) = 0
\end{equation}
and that the difference between the value of this quantity at times
$t + \Delta t/2$ and $t - \Delta t/2$ will be
\begin{equation}
- (\Delta t)^2 \sum_i ( DF_i^a(x,t) - DF_i^a(x-i,t) )
\end{equation}
which vanishes identically, because each plaquette which appears in
the evaluation of one term appears with opposite sign in the evaluation
of another term, as before.  Hence the Gauss constraint will be preserved
identically by the evolution.  

Eq. (\ref{updaterule}) does not give the Lie algebra content of the
electric fields, $(1/2) {\rm Tr} -i \tau^a E_i(x)$,
directly, but gives the result of a nondiagonal, nonlinear transformation
on them; to implement the update it is necessary to invert this 
transformation.  The inversion problem is not 
too bad (for $\Delta t \ll 1$), because the
expression on the lefthand side of Eq. (\ref{updaterule}) is close to
diagonal in the $E$ fields and the inversion can be performed iteratively, ie
\begin{eqnarray}
&& \frac{7}{6} \left( \frac{1}{2} {\rm Tr} -i \tau^a E_i(x) \right)
=  ({\rm r. \: h. \: s. \: of \; Eq. ( \protect{\ref{updaterule}})} ) + 
\\
&& \frac{1}{12} \left( 
\frac{1}{2} {\rm Tr} -i \tau^a E_i(x) E_i(x + i) +
\frac{1}{2} {\rm Tr} -i \tau^a E_i(x-i) E_i(x) 
- 2 \frac{1}{2} {\rm Tr} -i \tau^a E_i(x) \right) 
\nonumber
\end{eqnarray}
can be solved by making a guess for $E$, substituting into the 
righthand side, and getting a better guess for $E$.  I have written
the formula so that the righthand side depends almost only on $E$ at
the points 1 space ahead and behind of the point of interest, so the
convergence rate is doubled by alternately updating even and odd lattice
points.

I have implemented this algorithm numerically.  At single precision, choosing
the initial guess for $E_i(x,t + \Delta t/2)$ to be
$E_i(x, t - \Delta t/2)$, it is only necessary to iterate over the
lattice 4 times before the remaining correction comes on order roundoff
error.  If the inversion were conducted completely to infinite precision
then the algorithm would not gain or lose energy in the mean, because
the update rule is time reversal invariant and only depends on the minimal
amount of state information.  For a single precision implementation, 
because the iteration converges from above, the roundoff errors 
systematically boost the
energy, so that it rises by about 2 parts in $10^{8}$ per time step; but
this level is small enough to be of no concern if the total length of 
a simulation is kept under $t \simeq 5000$.  By going to double precision
and making the timestep very small to reduce the fluctuations in 
system energy I was able to verify that the algorithm does preserve 
energy in the mean.  I also observe that violation of the Gauss constraint
is only generated by roundoff error, as in the evolution with the
unimproved action.  As a final check I 
verified that the implementation produces
the expected dispersion relations when the initial state is a weak
sinusoidal excitation in one Lie algebra direction.

\subsection{Magnetic field and $N_{CS}$}

The next step in the program is to find the appropriate definition
of $dN_{CS}/dt$ for the improved Hamiltonian.  The approach is similar
to the unimproved case, where $\int (dN_{CS}/dt) dt$ is defined as 
\begin{equation}
\sum_{x,t} \sum_i \frac{1}{2 \pi^2} \left( \frac{1}{4}
\sum_{4 \Box_{jk}} \frac{1}{2} {\rm Tr} -i \tau^a U_{\Box_{jk}} \right)
\times \left( \frac{1}{4}
\sum_{4 \Box_{0i}} \frac{1}{2} {\rm Tr} -i \tau^a U_{\Box_{0i}} \right) \, ,
\end{equation}
where the first sum is over all points in the 4 dimensional grid, the
second is an average over the 4 plaquettes in the $jk$ plane ($ijk$ a
cyclic permutation) which touch the point, and the last is a sum over the
4 plaquettes which proceed one step forward or backward in the time 
direction and a step forward or backward in the $i$ direction from the point.
That is, it is an average over the electric fields leading into and
out of the point, taken a half timestep before and after the spacetime
point.  In the previous section I rearranged this sum into a sum over
the electric fields of a contribution coming from each end of the link 
on which the electric field resides, which is how the sum over 8
square plaquettes arose.  

To modify this definition so as to remove dimension 6 contamination, I
need to replace the sum over square plaquettes with a sum on
square and $1 \times 2$ rectangular plaquettes.  Again, because the 
plaquettes are planar and the improvement need only correct dimension 6
errors (dimension 4 in the electric or magnetic field), terms cubic in
the field strength will not arise and I need only consider
weak sinusoidal fields in one Lie algebra direction.
In a small $k$ expansion, the choice of spatial plaquettes should not
contain any $k^3$ term.

The contribution from the 4 square plaquettes around a point $x$
and in the $1,2$ plane for $U_i(x) = 1 + \epsilon i \tau^a \sin 
(k \cdot x + k_i/2)$ is
\begin{equation}
4 c_1 c_2 ( E_2 s_1 - E_1 s_2) \epsilon^2 \sin ( k \cdot x ) \, ,
\end{equation}
where $c_i = \cos k_i/2$.  The contribution from the 4 rectangular
plaquettes which have midpoints at $x$ is twice this,
and the choice of whether to use square plaquettes or midpoints of 
rectangular ones is ours (though it may become forced when improvement
is carried up to higher order).  The contribution from the
8 rectangular plaquettes which have a corner at $x$
is
\begin{equation}
4 c_1 c_2 ( 4 - s_1^2 - s_2^2) ( E_2 s_1 - E_1 s_2) \epsilon^2 
\sin ( k \cdot x )  \, .
\end{equation}
The correct combination should produce no $k^3$ term.
The requirement is solved by using $5/3$ the sum over square plaquettes
(or $5/6$ the sum over rectangular ones with midpoints at $x$)
plus $-1/6$ the sum over rectangular plaquettes with corners
at $x$.  For the electric field it is not necessary
to use rectangular plaquettes which stretch two units in the time direction,
because the field variation in this direction is suppressed by a factor
of $(\Delta t)^2$,
and the right combination at the point $x,t$ (making parallel transports
using the links at time $t$) is
\begin{eqnarray}
& & \frac{4}{3} \left( \frac{1}{2} {\rm Tr} -i \tau^a E_i(x)
+  \frac{1}{2} {\rm Tr} -i \tau^a E_i(x-i) \right)(t+\Delta t/2) -
\\ & &
\frac{1}{6} \left( 
 \frac{1}{2} {\rm Tr} -i \tau^a E_i(x) E_i(x+i) +
 \frac{1}{2} {\rm Tr} -i \tau^a E_i(x-2i) E_i(x-i) \right)(t+\Delta t /2) +
\nonumber \\ & &
 \frac{4}{3} \left( \frac{1}{2} {\rm Tr} -i \tau^a E_i(x)
+  \frac{1}{2} {\rm Tr} -i \tau^a E_i(x-i) \right)(t-\Delta t/2) -
\nonumber \\ & &
\frac{1}{6} \left( 
 \frac{1}{2} {\rm Tr} -i \tau^a E_i(x+i) E_i(x) +
 \frac{1}{2} {\rm Tr} -i \tau^a E_i(x-i) E_i(x-2i) \right)(t-\Delta t/2) \, .
\nonumber
\end{eqnarray}

This differs only by terms of order $(\Delta t)^2$ from the same 
expression with the trace taken over each $E$ in the two $E$ terms, rather
than over the product.  This is because the order $\Delta t$ correction
arises from a commutator term which 
enters with opposite sign in the $t-\Delta t/2$ and $t+\Delta t/2$
contributions.
Since there are already inevitable $O( (\Delta  t)^2)$ errors
in the evolution of the system, I am free to choose
the more convenient definition without worrying about the order 
$(\Delta t)^2$ change it will make in the results.
Choosing to trace each $E$ separately, 
I can rearrange the definition to be
\begin{eqnarray}
2 \pi^2 \Delta N_{CS} & = & \sum_{x,i} B_i^a(x,t)
\frac{1}{2} \left(  \frac{1}{2} {\rm Tr} -i \tau^a E_i(x, t+\Delta t/2 ) +
\right. \nonumber \\ & & \left. \qquad \qquad \qquad
 \frac{1}{2} {\rm Tr} -i \tau^a E_i(x,t - \Delta t/2) \right)
\\
B_i^a(x) & \equiv & \frac{-1}{12} F_{jk}^a(x+2i) + \frac{7}{12}
F_{jk}^a (x + i)
+ \frac{7}{12} F_{jk}^a (x) 
- \frac{1}{12} F_{jk}^a ( x - i )
\\
4 F_{jk}^a(x) & \equiv & \frac{5}{3} \sum_{4 \Box_{ij}} \frac{1}{2} 
{\rm Tr} -i \tau^a U_{\Box_{jk}}
- \frac{1}{6} \sum_{8 \sqsubset \! \sqsupset_{jk}}
\frac{1}{2} {\rm Tr} -i \tau^a U_{\sqsubset \! \sqsupset_{jk}}
\end{eqnarray}
where all definitions should be clear from earlier usage and 
parallel transports of $F_{jk}^a$ are implied along the straight
line paths to the point $x$.

This gives a usable definition for the evolution of $N_{CS}$ which is
free of dimension 6 contamination.

\subsection{Chemical potential for $N_{CS}$ }

Now I should apply a chemical potential for $N_{CS}$ in order to study
its motion due to infrared physics.  In analogy with the unimproved
Hamiltonian case, the chemical potential term should modify the evolution
of the $E$ fields in a way which makes the overall energy change
by $- \mu N_{CS}$.  In the unimproved case this was accomplished by making
\begin{equation}
\delta E_{\alpha} = - \frac{\mu \delta t}{2 \pi^2} B_{\alpha} \, ,
\end{equation}
 so the change in the
electric energy was 
\begin{equation}
E_{\alpha} \delta E_{\alpha} = - \frac{\mu \delta t E_{\alpha}
B_{\alpha}}{2 \pi^2} = - \mu \delta N_{CS}  \, .
\end{equation}  
Then I modified the definition so that the 
linear combination of $E$ fields representing a Gauss constraint violation
were not excited after all, by orthogonally projecting to the 
constraint surface.  This did not change the energy because
these modes had started out zero anyway.

In the improved case, it is easiest to begin by thinking about the 
continuum time system.  The change in $N_{CS}$ is 
\begin{equation}
dN_{CS} = \frac{E_{\alpha} B_{\alpha} dt}{ 2 \pi^2} \, ,
\end{equation} 
but the shift in the energy is
$\delta E_{\alpha} M_{\alpha \beta} E_{\beta}$, so $\delta E$ should be
\begin{equation}
\label{BshiftsE}
\delta E = \frac{ (M^{-1})_{\alpha \beta} B_{\beta} dt}{ 2 \pi^2 } \, ,
\end{equation}
before worrying about 
preserving the Gauss constraint.  Note that the evolution equation
for $E$ already includes a term like $- M^{-1}_{\alpha \beta} DF_{\beta} dt$,
so the magnetic field appears in the same place as $DF_i^a(x)$.  This
describes a possible implementation in the discrete case as well,
namely that Eq. (\ref{updaterule}) is modified by replacing
$- DF_i^a(x)$ by $-DF_i^a(x) - \mu B_i^a(x)/2 \pi^2$.

Another way of arriving at this result is to note that the Hamiltonian
defines a metric on the space of $E$, and that in the improved
Hamiltonian the natural basis
$E_{\alpha}$ is no longer orthogonal.  Rather, the orthogonal basis
is $(M^{1/2})_{\alpha \beta} E_{\beta}$.  In terms of this basis, the
definition of $N_{CS}$ is
\begin{equation}
\frac{dN_{CS}}{dt} = \frac{ ( (M^{\frac{1}{2}})_{\alpha \beta} E_{\beta} )
(M^{-\frac{1}{2}})_{\alpha \gamma} B_{\gamma} }{2 \pi^2} \, ,
\end{equation}
and so the change in the electric fields should be
\begin{equation}
\delta ( (M^{\frac{1}{2}})_{\alpha \beta} E_{\beta}) = 
\frac{ (M)^{- \frac{1}{2}}_{\alpha \beta} B_{\beta} \delta t}{2 \pi^2} \, ,
\end{equation}
which leads to Eq. (\ref{BshiftsE}).

Now although the new definition of the magnetic field should be free
of the leading order of operators responsible for it not satisfying the
Bianchi identity, so the satisfaction of the Gauss constraint 
should be better in the infrared,
the ultraviolet violation of the Gauss law is just as bad as in the 
unimproved case, and it is again necessary to change the definition of
the action of a chemical potential to prevent the violation of the
Gauss constraint.  In analogy with the unimproved case, the correct
modification is to orthogonally project back to the constraint surface.
Again, orthogonal should mean with respect to the metric defined by $H_E$,
so the most general motion orthogonal to the constraint surface is
\begin{equation}
\label{orthogonalmotion}
\delta E_{\alpha} = \kappa_{\beta} c_{\beta \alpha} \qquad {\rm not}
\qquad \delta E_{\alpha} = 
\kappa_{\beta} c_{\beta \gamma} M_{\gamma \alpha} \, .
\end{equation}

For small $\kappa$, the former changes the system energy by
\begin{equation}
\delta H_E \simeq E_{\alpha} M_{\alpha \beta} \delta E_{\beta} = 
E_{\alpha} M_{\alpha \beta} c_{\gamma \beta} \kappa_{\beta} \, ,
\end{equation}
which is suppressed by a power of the adherence to the Gauss constraint,
whereas the latter changes the system energy by
\begin{equation}
E_{\alpha} M_{\alpha \beta} M_{\beta \epsilon}  c_{\gamma \epsilon} 
\kappa_{\beta} \, ,
\end{equation}
which is nonzero even if the Gauss constraint is satisfied; so the 
modification to $E$ is not orthogonal to the constraint surface.

A nice way to understand this conclusion is to remember 
that $M^{1/2}E$ are orthogonal; the Gauss constraint is
\begin{equation}
G_{\alpha} = (c_{\alpha \beta} (M^{\frac{1}{2}})_{\beta \gamma})
((M^{\frac{1}{2}})_{\gamma \delta} E_{\delta}) = 0 \, ,
\end{equation}
so the most general correction to the electric field which is orthogonal
to the constraint surface should be
\begin{equation}
\delta ( (M^{\frac{1}{2}})_{\alpha \beta} E_{\beta} ) = 
\kappa_{\gamma} (M^{\frac{1}{2}})_{\alpha \epsilon} c_{\gamma \epsilon}
\qquad {\rm or} \qquad \delta E_{\alpha} = \kappa_{\beta} c_{\beta \alpha}
\label{setA0zero}
\end{equation}
which is the same as Eq. (\ref{orthogonalmotion}).

I can now find an approximate projection algorithm in strict analogy with
the unimproved Hamiltonian case.  A step in a dissipative or quench algorithm
for the orthogonal projection is
\begin{equation}
\delta E_{\alpha} = \kappa_{\beta} c_{\beta \alpha} \, , \qquad
\kappa_{\beta} = - \gamma c_{\beta \gamma} M_{\gamma \epsilon} E_{\epsilon}
\end{equation}
where stability now demands $\gamma < 1/8$.  This step is orthogonal
to the constraint surface, and repeated application is guaranteed to be
a projection; a single application at each step is enough to keep the 
departure from the constraint surface under control.  A better algorithm,
again in analogy with the unimproved system, is to make an initial correction
with $\kappa$ equal to $m<1$ times the total $\kappa$ of 
the previous step, and then to 
apply the dissipative algorithm twice, with different stepsizes.

Note that Eq. (\ref{orthogonalmotion}) is of exactly the same form as
in the unimproved case.
To understand why, recall the origin of the Gauss constraint; it is the
Lagrange equation for the $A_0$ field (or the time direction link matrix),
and when it is not satisfied it means we have inadvertently not set
$A_0$ to zero; since the electric field only corresponds to the time 
derivative of the link matrix when $A_0$ is zero, we should restore $A_0$
to zero, and Eq. (\ref{setA0zero}) is precisely how the system changes
as we modify $A_0$ by $\kappa$.

This latter observation explains how to project to the constraint surface
in the discrete time case.  The modification of the electric fields
should take the form of a time dependent gauge transformation at each
point, with magnitude chosen to restore the Gauss constraint everywhere.
In other words the correct orthogonal projection to the constraint 
surface is a choice of a group element $g(x)$ at each point $x$ which
modifies the $E$ matricies through
\begin{equation}
E_i(x, t + \Delta t/2) \rightarrow g(x,t) E_i(x,t + \Delta t/2) 
 U_i(x,t) g^{\dagger}(x + i,t) U^{\dagger}_i(x,t)  \, .
\end{equation}
The $g(x)$ should be chosen such that 
\begin{eqnarray}
G^a(x,t) & = & 
\sum_i \left[ \frac{4}{3} \left( \frac{1}{2}{\rm Tr} 
-i \tau^a E_i(x,t + \Delta t/2)
- \frac{1}{2}{\rm Tr} -i \tau^a E_i(x-i,t + \Delta t/2) \right)
- \right. 
\nonumber \\ &&
\frac{1}{12} \left( \frac{1}{2}{\rm Tr} -i \tau^a E_i(x,t + \Delta t/2) 
E_i(x+i, t + \Delta t/2)
- \right.
\nonumber \\ & & \left. \left. \qquad
\frac{1}{2}{\rm Tr} -i \tau^a E_i(x-2i,t + \Delta t/2) 
E_i(x-i,t + \Delta t/2 ) 
\right) \right] 
\end{eqnarray}
(parallel transports at time $t$ implied) is zero.  Again the job of
finding the right $g$ is intractable, but a reasonable guess (the basis
of the dissipative algorithm) is 
\begin{equation}
\frac{1}{2} {\rm Tr} -i \tau^a g(x,t) = - \gamma G^a(x,t) \, ,
\end{equation}
and a better guess is to use $m$ times the Lie algebra element used in
the previous time step, and then to apply the above dissipative step
twice with different coefficients $\gamma$.  The stability condition
is now $ -1/3 < (1 - \omega \gamma_1)(1 - \omega \gamma_2) < 1$
for all $0 < \omega < 16$, and I find that
a good choice is $m \sim 1 - 1.6 \Delta t$, $\gamma_1 = 5/64$, $\gamma_2
= 5/32$.  Using these values, the Gauss constraint is as nearly 
satisfied as in the unimproved Hamiltonian, and the relation
$\Delta H = - \mu \Delta N_{CS}$ is satisfied to $<1\%$ (after energy
gain due to roundoff errors are taken into account; I checked this in
a double precision implementation, where there is no energy gain from
roundoff error), which should assure that the systematic errors in the
determination of the linear response to a chemical potential, due to
not using an exact orthogonal projection, 
are of the same size \cite{paper1}.

This provides a satisfactory implementation of chemical potential in
the improved Hamiltonian system.

\subsection{Thermalization}

Next I extend the thermalization algorithm developed for the unimproved
Hamiltonian to the improved one.  I will begin by showing how to thermalize
the system in the continuum time case.  The technique does not strictly
work for finite $\Delta t$, but the failures are of order $(\Delta t)^2$, 
and since the time evolution and the definition of $N_{CS}$
already introduce systematic errors of at least this size I will ignore
this problem and find a discrete implementation which is correct up to
these stepsize ambiguities.

In the continuum time case, I want to thermalize the system with Hamiltonian
given in Eq. (\ref{rightHamiltonian}), subject to the Gauss constraints 
given in Eq. (\ref{rightGaussconstraint}).  The subspace of momenta at
fixed $U$ is an inner product space with $2 H_{E}$ the inner product,
and the Gauss constraints are linear, so I can implement the thermalization
discussed in Subsec. \ref{unimp_thermalization}, modified only to account
for the new inner product; in other words I need only find a thermalization
algorithm for the fixed $U$ electric field part of the Hamiltonian and
then use the equations of motion to stir the thermalization between electric
and magnetic modes.  

If we had an orthonormal
basis which partitioned into Gauss constraints and dynamical degrees of
freedom, we would choose the dynamical degrees of freedom to be Gaussian
distributed and the constraints to be zero, and the electric fields would
be thermalized.  Equivalently I could Gaussian distribute both of them and
then orthogonally project to the constraint surface.  The continuum time
orthogonal projection algorithm is given already in the previous section,
and a Gaussian distribution in any orthonormal basis is Gaussian in any
other, so the only remaining problem is to find an orthonormal basis.
In the previous section we saw that the basis $M^{1/2}E$ is orthonormal.
Therefore, a thermalization of the electric fields consists of choosing
the $M^{1/2}E$ Gaussian, inverting $M^{1/2}$ to find $E_{\alpha}$, and
applying the orthogonal projection by repeated application of the 
continuum time dissipative algorithm of the last section.  

The inversion can be carried out as follows: note that $M$ is $(7/6) I$ plus
a small off diagonal term, $M_{\alpha \beta} = 7/6( \delta_{\alpha \beta} 
+ m_{\alpha \beta})$; write
\begin{eqnarray}
E_{\alpha} & = & ( M^{-\frac{1}{2}})_{\alpha \beta} (M^{\frac{1}{2}}E)_\beta
\\
& = & ( (\frac{7}{6} (I + m))^{- \frac{1}{2}})_{\alpha \beta} 
 (M^{\frac{1}{2}}E)_\beta
\nonumber \\
& = & \sqrt{\frac{7}{6}} \left( \delta_{\alpha \beta} - \frac{1}{2} m_{\alpha
\beta} + \frac{3}{8} m^2_{\alpha \beta} - \frac{5}{16} m^3_{\alpha \beta}
+ \ldots \right)  (M^{\frac{1}{2}}E)_\beta \, ,
\nonumber 
\end{eqnarray}
in other words, expand $M^{-1/2}$ in a Taylor series.  The absolute
magnitude of the largest
possible eigenvalue of $m$ is 1/7, so the Taylor series converges rather
well; and the inefficiency of the algorithm does not matter because it
is only called occasionally, with several (order 100) lattice updates 
inbetween.

It is not completely trivial to carry this algorithm over directly to the
discrete time setting.  For one thing it is not clear that randomizing
$E_i(x , t + \Delta t/2)$ is the appropriate thing to do.  If we did, the
thermalization algorithm would not be quite time reversal invariant,
as ${\rm Tr} - i \tau^a E_i(x , t + \Delta t/2 )$ 
would then be uncorrelated with $DF_i^a(x,t)$,
but backwards evolving, we would find that 
 ${\rm Tr} - i \tau^a E_i(x , t - \Delta t/2 )$ {\it was} correlated
with $DF_i^a(x,t)$.  Also, the $E$ are matricies in SU(2), not Lie 
algebra elements, and while there is a mapping from SU(2) to the Lie
algebra, given by $1/2 {\rm Tr} -i \tau^a$, which is one to one from
half of SU(2) and not difficult to invert in this patch, the
determinant of the map is not 1, ie it does not preserve the measure 
of SU(2).  Also, the energy depends not on $1/2$  
the square of the Lie algebra element,
but on $1 - 1/2 {\rm Tr}$ of the group element, which, in terms of 
$E_i^a \equiv 1/2 {\rm Tr} -i \tau^a E$, is $1 - \sqrt{ 1 - E^a E^a}
\neq E^a E^a/2$.  In terms of the Lie algebra element, nonquadratic
terms will appear in the Hamiltonian, both from the measure and from
the way the group element is traced to get the energy.  Therefore the
thermalization algorithm I have used is, strictly speaking, wrong.

However, I can define a Lie algebra element ``electric field'' 
at integer time steps, for which all of these ambiguities would
appear as order $(\Delta t)^2$ or $(\Delta t)^2 / \beta_L$ corrections,
and can in practice be ignored.  (It is not even clear that there is a
meaning for temperature beyond leading order in $(\Delta t)^2$ 
in the discrete time system.)  The definition is
\begin{eqnarray}
\frac{7}{6} E_i^a(x,t) \! & - \! & 
\frac{1}{12} ( E_i^a(x-i,t) + E_i^a(x+i,t)) =
\\ & & 
\frac{1}{\Delta t} \left[ \frac{- (\Delta t)^2}{2} DF_i^a(x) +
\frac{4}{3} \left( \frac{1}{2} {\rm Tr} - i \tau^a E_i(x) \right)
(t - \Delta t/2) - \right.
\nonumber  \\ & & \left.
\frac{1}{12} \left(  \frac{1}{2} {\rm Tr} - i \tau^a E_i(x) E_i(x-i) +
 \frac{1}{2} {\rm Tr} - i \tau^a E_i(x+i) E_i(x) \right) ( t - \Delta t/2)
\right] 
\nonumber 
\end{eqnarray}
(all parallel transports to be conducted with respect to the links $U_i$
at time $t$),
from which it is simple to continue the time evolution of the system.  The
Gauss constraints on either side of time $t$ will be satisfied provided
that $E_i^a(x,t)$ obeys
\begin{eqnarray}
\label{funnyGauss}
0 & = & \sum_i\left[ \frac{5}{4} \Big( E_i^a(x,t) - E_i^a(x-i,t) \Big)
- \frac{1}{12} \Big( E_i^a(x+i ,t) - E_i^a(x-2i,t) \Big) \right] 
\\
& = & c_{\alpha \beta} M_{\beta \gamma} E_{\gamma} \, . \nonumber 
\end{eqnarray}
I will thermalize this $E_i^a$ by drawing it from 
the thermal distribution of the Hamiltonian
$E_{\alpha} M_{\alpha \beta} E_{\beta}/2$.  The algorithm is precisely the
continuum time algorithm presented above.  

\begin{table}
\begin{tabular}{|c|c|c|c|} \hline
 quantity &  at $\Delta t = 0.05$ &  at $\Delta t = 0.20$ & 
  at $\Delta t = 0.30$ \\ \hline \hline
$H_{U \; {\rm improved}} \beta_L / ( 3 N^3) $ & $1.031 \pm .0016 $
& $1.1081 \pm .0015$  & $1.2278 \pm .0016 $ \\ \hline
$1 \times 1$ Wilson loop &  $ \; .84865 \pm .00024 \; $
&  $ \; .83774 \pm .00021 \; $ & $ \; .82087 \pm .00025 \; $  \\ \hline
$1 \times 2$ Wilson loop &  $.72376 \pm .00035 $
&  $.70763 \pm .00034 $ & $.68274 \pm .00038 $ \\ \hline
$1 \times 3$ Wilson loop &  $.6201 \pm .0005$
&  $.6005  \pm .0005  $ & $.5703 \pm .0006 $  \\ \hline
$1 \times 4$ Wilson loop &  $.5316 \pm .0007 $ 
&  $.5104  \pm .0007  $ & $.4768  \pm .0007  $  \\ \hline
$1 \times 5$ Wilson loop &  $.4560 \pm .0008 $
&  $.4338  \pm .0008  $ & $.3988  \pm .0008  $ \\ \hline
$2 \times 2$ Wilson loop &  $.5412 \pm .0010 $
&  $.5222  \pm .0010  $ & $.4909  \pm .0010  $  \\ \hline
$2 \times 3$ Wilson loop &  $.4121 \pm .0009 $
&  $.3925  \pm .0008  $ & $.3592  \pm .0009  $  \\ \hline
$2 \times 4$ Wilson loop &  $.3149 \pm .0010 $
&  $.2967  \pm .0009  $ & $.2644  \pm .0010  $  \\ \hline
$2 \times 5$ Wilson loop &  $.2410 \pm .0010 $
&  $.2246  \pm .0010  $ & $.1949  \pm .0010  $  \\ \hline
$3 \times 3$ Wilson loop &  $.2843 \pm .0015 $
&  $.2660  \pm .0015  $ & $.2344  \pm .0015  $  \\ \hline
$3 \times 4$ Wilson loop &  $.1975 \pm .0011 $
&  $.1824  \pm .0010  $ & $.1548  \pm .0010  $  \\ \hline
$3 \times 5$ Wilson loop &  $.1377 \pm .0011 $
&  $.1255  \pm .0011  $ & $.1018  \pm .0010  $  \\ \hline
$4 \times 4$ Wilson loop &  $.1251 \pm .0015 $
&  $.1143  \pm .0015  $ & $.0924  \pm .0015  $  \\ \hline
$4 \times 5$ Wilson loop &  $.0848 \pm .0011 $
&  $.0722  \pm .0011  $ & $.0546  \pm .0010  $  \\ \hline
$5 \times 5$ Wilson loop &  $.0467 \pm .0015 $
&  $.0424  \pm .0015  $ & $.0286  \pm .0015  $  \\ \hline
\end{tabular}
\caption{ \label{thermstepsize} Stepsize dependence of thermodynamic
properties in the thermalization algorithm for the improved Hamiltonian
lattice, for a $16^3$ lattice at lattice inverse temperature $\beta_L = 5$.}
\end{table}

To check that this thermalization algorithm is reasonable, in the sense
that it gives thermodynamic properties which change only very weakly with
stepsize when the stepsize is small, I have evaluated
the magnetic energy and Wilson loops at different stepsizes.
Some results for a $16^3$ lattice at lattice temperature $\beta_L = 5$ are
presented in Table \ref{thermstepsize}.  For each column I have used the
algorithm to thermalize an initial configuration and then continued to
implement the algorithm for 80 $E$ field randomizations, recording each 
observable once each $E$ field randomization.  The error bars are from
statistics, corrected for correlations because one $E$ field randomization
is insufficient to fully rethermalize the system.  There are clear stepsize
effects which are consistent with the expected $(\Delta t)^2$ behavior.
The changes are significant for the two larger stepsizes in the table, so
$\Delta t$ cannot be taken to be this large; but the zero $\Delta t$ 
extrapolated magnetic energy differs from the
$\Delta t = 0.05$ value by only $0.5 \%$, which
will lead to a systematic error in Lyapunov exponents of order $0.5 \%$ and
in the motion of $N_{CS}$ under chemical potential of order $1.5 \%$,
which will be less than statistical measuring error.  $\Delta t = 0.05$ is
therefore sufficiently small that the remaining stepsize systematics 
can be neglected.  Note also that these
systematics will apply to all measurements, so they will not affect
any conclusions about the existence of a fine lattice spacing limit.

Based on this analysis, and because the evolution of the system will have
stepsize errors of similar size, I have used $\Delta t = 0.05$ 
everywhere in what follows, unless noted otherwise. 

\section{Numerical results}
\label{Results}

I have implemented the improved Hamiltonian system developed in the
previous section numerically, and used it to study the motion of
$N_{CS}$ under a chemical potential.  It has become conventional in the
field to quote the rate of $N_{CS}$ violation in the symmetric phase
in terms of a dimensionless quantity $\kappa$.  The linear response
coefficient to a chemical potential is
\begin{equation}
\Gamma_\mu \equiv \frac{ T \langle \dot{N}_{CS}\rangle}{\mu V} 
= \kappa_{\mu} (\alpha_W T)^4 \, ,
\end{equation}
which in lattice units is
\begin{equation}
\kappa_{\mu} = \frac{ (\beta_L \pi)^4 \Delta N_{CS} }{\beta_L \mu N^3 t} \, .
\end{equation} 
By a detailed balance argument \cite{RubakShap2,paper1} this is half the
rate of $N_{CS}$ diffusion,
\begin{equation}
\Gamma_d \equiv \lim_{t \rightarrow \infty}  
\frac{ \langle ( N_{CS}(t) - N_{CS}(0) )^2 \rangle }{ V t } = 
\kappa_d (\alpha_W T)^4 \, , \qquad \kappa_d = 2 \kappa_\mu
\label{diffrate}
\end{equation}
measured in \cite{AmbKras}.  

\begin{table}
\begin{tabular}{|c|c|c|c|} \hline
$\beta_L$ & $N$ & number of runs & $\kappa_{\mu}$ \\ \hline \hline
3 & 16 & 3  & $.181 \pm .012 $ \\ \hline
4 & 16 & 6  & $.310  \pm .017  $ \\ \hline
5 & 16 & 16 & $.389 \pm .021 $ \\ \hline
6 & 16 & 27 & $.413 \pm .024 $ \\ \hline
8 & 24 & 16 &  $.487 \pm .033 $ \\ \hline
10 & 30 & 18 & $ .537 \pm .036 $ \\ \hline 
\end{tabular}
\caption{\label{NCStable} Rate of motion of $N_{CS}$ under a chemical
potential at several lattice spacings.}
\end{table}

I have evaluated the rate of $N_{CS}$ motion under a chemical 
potential for several lattice spacings (lattice reciprocal temperatures),
using in every case a chemical potential $\mu = 2 / \beta_L$, which
previous work shows is easily small enough to fall in the linear 
response regime \cite{paper1}.  I used lattice sizes such that 
$N \geq 3 \beta_L$ 
in most cases, because it has been shown that finite lattice size
effects vanish already at $N = 2 \beta_L$ \cite{AmbKras}.  The
evolutions were each of length $t = 400$ in lattice units; 
for each lattice coarseness I used several
such evolutions from independent initial configurations drawn from
the thermal ensamble by the thermalization algorithm.
The error bars are statistical; when there were several
runs they were determined by the fluctuations between independent runs.
I also computed them by assuming the statistical error comes from Brownian
motion in $N_{CS}$, with a rate given by Eq. (\ref{diffrate}); the
two estimates agree within error, so I used the latter estimate when
there were few independent runs.  

\begin{figure}
\centerline{\psfig{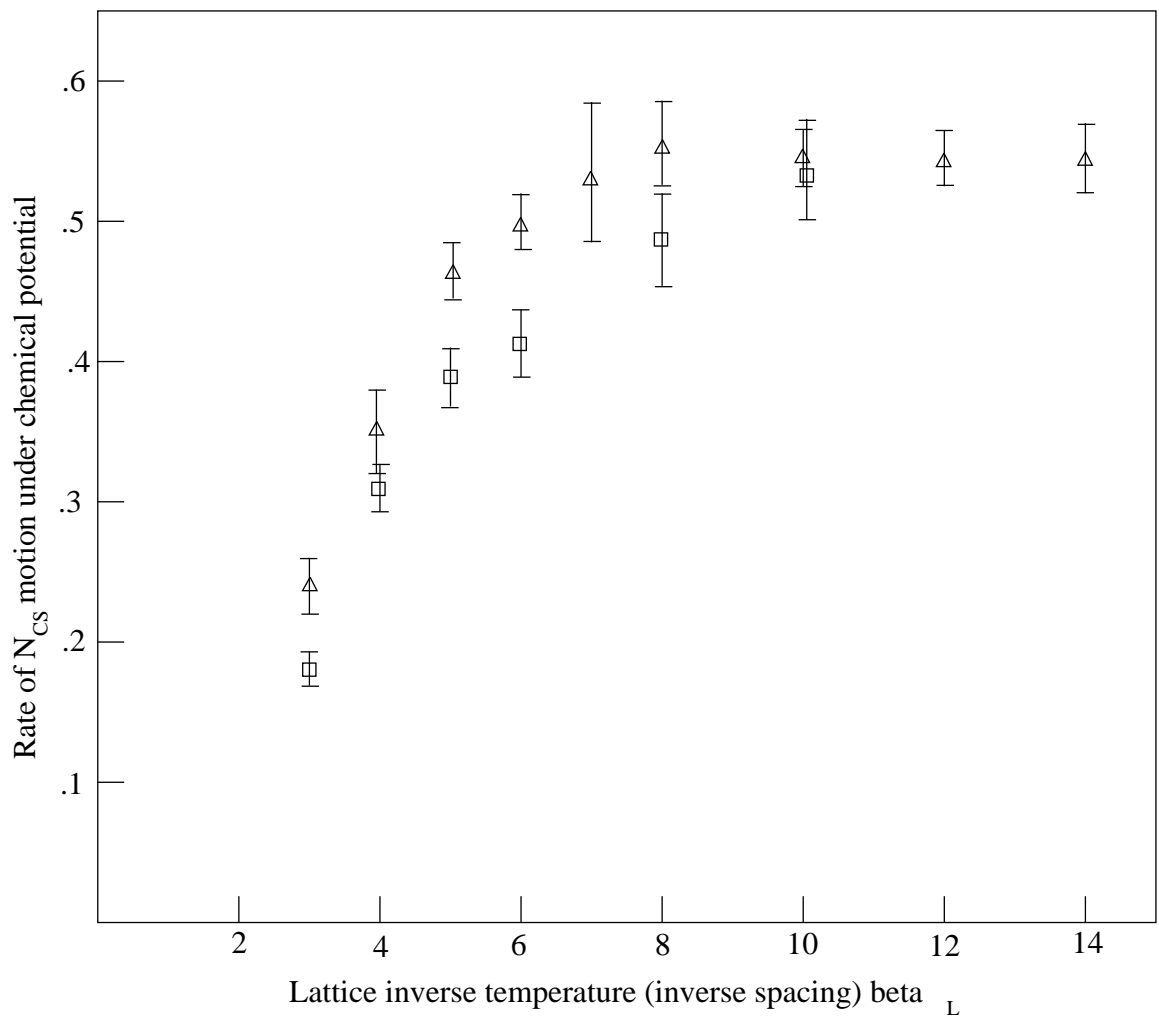}}
\caption{Rate of motion of $N_{CS}$ under a chemical potential for
several lattice inverse 
temperatures (reciprocal lattice spacings).  The squares are
the data using the improved Hamiltonian, the triangles are data using the
Kogut Susskind Hamiltonian, taken from \protect{\cite{AmbKras,paper1}}.
\label{NCSfigure}}
\end{figure}

My results are presented in Table \ref{NCStable} and plotted, along with 
some results of \cite{AmbKras,paper1} for the unimproved Hamiltonian,
in Figure \ref{NCSfigure}.  
As the lattice becomes finer, $\kappa_{\mu}$ appears to converge to
a lattice spacing independent result.  The rate of 
convergence is slightly worse 
for the improved Hamiltonian than for the unimproved one, so
the physics which establishes the rate of $N_{CS}$ diffusion
depends on the Debye screening length being well shorter than the length 
scale of nonperturbative physics.  In fact, from the data presented it
is not entirely clear that the improved lattice results have converged
to the fine lattice spacing limit; but the numerical demands rise as the 4'th
power of $\beta_L$, and the improved Hamiltonian requires 5 times as many
flops to evolve the same lattice 4-volume as the KS Hamiltonian ,
so I was unable to get useful data on finer lattices than those presented.
I will assume here that the answer has converged at $\beta_L = 10$, since even
if the convergence rate is Debye mass limited, then the convergence should
be as good here as in the unimproved system at $\beta_L \simeq 8$. 

For the current level of statistics the results of the improved system
are compatible with those found using the Kogut-Susskind Hamiltonian.
However, the still substantial error bars leave open the possibility that
the rates differ by order $5 \%$ to $ 10 \% $.  To compare the rates
to $1 \%$ would require approximately 50 times the integration time
used here, or about $2 \times 10^{15}$ floating point operations, which
is prohibitive.

To examine whether the two Hamiltonians really do produce different 
infrared dynamics,
or whether I just have insufficiently refined lattices and insufficient 
statistics, I have measured another quantity, the maximal Lyapunov
exponent, for the improved Hamiltonian system.  The maximal Lyapunov
exponent is defined as the rate at which two almost identical initial
configurations diverge, as measured by some gauge invariant phase space
metric.  It is a better quantity to use to compare the improved and
unimproved Hamiltonian, because it depends on less equipment than the
measurement of $\kappa_{\mu}$, requiring only the evolution and thermalization
algorithms, and because the statistics improve more quickly, rising 
roughly as $ \sqrt{N^3 t/ \beta_L^4}$, whereas for $\kappa_{\mu}$ the
statistics improved as 
$\mu \sqrt{ \kappa_{\mu} N^3 t / (\beta_L \pi)^4}/2$.
It will therefore be possible to compare the maximal Lyapunov exponents
of the two theories to order $1 \% $ without excessive numerical demands.

\begin{table}
\begin{tabular}{|c|c|c|c|c|c|} \hline
$\beta_L$ & $N$ & improved $ \lambda_{max} $ & 
KS $ \lambda_{max} $ & improved $\beta_L \lambda_{max}$
& KS $\beta_L \lambda_{max}$ \\ \hline \hline
2 & 12 &  & $.659 \pm .002$ &  & $1.318 \pm .004$ \\ \hline
2.5 & 12 & $.4932 \pm .005$  & $.586 \pm .002$ & $1.233 \pm .013 $
 & $1.465 \pm .005$ \\ \hline
3 & 12 & $.399 \pm .0034$ & $.497 \pm .004$ & $1.197 \pm .010$ & $1.491
\pm .012$ \\ \hline
4 & 12 & $.2863 \pm .0021$ & $.350 \pm .003$ & $1.145 \pm .008$ & 
$1.400 \pm .012$ \\ \hline
5 & 20 & $.2280 \pm .0018 $ & $.260 \pm .002$ & $1.140 \pm .009$ &
$1.300 \pm .010$ \\ \hline
6 & 20 & $.1908 \pm .0010 $ &  & $1.145 \pm .006$ &  \\ \hline
7 & 20 & $.1630 \pm .0027 $ & & $1.141 \pm .019$ &  \\ \hline
7.5 & 20 &  &  $.163 \pm .001$  &  & $1.222 \pm .008$ \\ \hline
8.75 & 20 &  & $.137 \pm .001$  &  & $1.199 \pm .009$ \\ \hline
10 & 20 &  &  $.120 \pm .001 $ &  &  $1.20 \pm .01 $ \\ \hline 
\end{tabular}
\caption{\label{Lyaptable} Lyapunov exponent for improved and Kogut-Susskind
Hamiltonian at several lattice spacings.  The improved data is new, the
KS data is taken from \protect{\cite{Krasnitz}}.
Each Hamiltonian system shows
a fine lattice limit, but the value of the limit differs by about $6 \%$ 
between the Hamiltonian systems.}
\end{table}

I will base my approach on that of Krasnitz \cite{Krasnitz}. I 
draw an initial condition from the thermal ensamble 
and then perturb it slightly;
then I allow the original configuration and the perturbed version to
evolve under the classical equations of motion, periodically measuring
the metric distance.  At first the separation is dominated by
transients; then there is an epoch of exponentially growing separation,
and finally the separation saturates.  It is the exponential epoch which
I use.  I repeat for several thermal initial configurations to determine
and improve the statistics.

In my implementation I start by getting a thermal initial condition by
applying my thermalization algorithm at single precision and a stepsize
of $\Delta t = 0.05$; I then convert to double precision, projecting the
fields to the $SU(2)$ manifold and quenching the Gauss constraint 
violations to the level of double precision roundoff error.  I then
copy this initial configuration and evolve it one time step, both with and
without a tiny ($10^{-13}$) chemical potential, to generate initial 
differences between configurations at the level of double precision 
roundoff error.  I let the configurations evolve, 
tracking the metric distances
\begin{equation}
D_E = \sum_{x,i} | \frac{1}{2} {\rm Tr} E_i(x) 
- \frac{1}{2} {\rm Tr} E_i'(x) | \, , \qquad
D_M = \sum_{\Box} | \frac{1}{2} {\rm Tr} U_{\Box} - \frac{1}{2}
{\rm Tr} U_{\Box}' | \, ,
\end{equation}
as proposed in \cite{Krasnitz}.  The choice of metric does not matter
much as long as it is gauge invariant; the two choices above give almost
exactly the same Lyapunov exponent for a run, and cannot be used as
independent measurements.  To eliminate transients 
I throw out all data until the metric distance
has grown by a factor of 500; I then follow the divergence of the 
two paths, stopping when the metric distance is within a factor of 500
of saturation.  This gives a very conservative fiducial period of evolution,
over which the metric distance grows by a factor of $10^9$.  (This is
the advantage of using double precision.)  I also use
lattice sizes well larger than $\beta_L$ to prevent finite size
effects.  To get the Lyapunov exponent I make a linear regression fit of
$\ln D_E$ or $\ln D_M$ against time for the data in this fiducial period.
To test the technique, I repeated one of the measurements of
Krasnitz, using the unimproved Hamiltonian and $\beta_L = 4$, $N = 12$;
my value of $ \lambda_{max} = .348 \pm .004$ is within
statistical error of the result presented in \cite{Krasnitz}.

The technique thus tested, I repeated it for several lattice spacings,
using the improved Hamiltonian; the results are presented in Table 
\ref{Lyaptable}, which also contains the results of Krasnitz 
\cite{Krasnitz} for comparison.  It is very clear from the table that
$\beta_L \lambda_{max}$ has a good fine lattice limit, both for the
improved and the unimproved Hamiltonian.  It reaches its limit much
faster in the improved case.  The limit for the improved Hamiltonian
differs from that for the KS Hamiltonian by about $5 \%$, significant at
about $5 \sigma$, confirming that while the fine lattice limit of the 
infrared dynamics exists within a given lattice regulation, its value 
depends on the specifics of that regulation.

As a final test of stepsize dependence, I recomputed the $\beta_L=4$
datapoint using $\Delta t = 0.1$ and $0.20$ (for both the 
thermalization algorithm and the evolution); the results for $\beta_L 
\lambda_{max}$ are $1.147 \pm .008$ and $1.200 \pm .010$, respectively.  
This agrees to within error
with a quadratic dependence in stepsize.  Back-extrapolating to zero
stepsize changes the results in the table by less than statistical error.

Note also that the limit of the Lyapunov exponent in both regulations
falls well below twice the damping rate of a plasmon at rest, which 
is \cite{Pisarski}
\begin{equation}
2 \frac{ 6.635 N_c g^2 T}{24 \pi} = .3520 g^2 T = 1.408 \frac{g^2 T}{4}
\, ,
\end{equation}
which would correspond to a value for $\beta_L \lambda_{max}$ of 
$1.408$.

\section{Conclusion}
\label{Conclusion}

An improved Hamiltonian can be constructed for the evolution of the 
classical Yang-Mills equations of motion on a lattice.  
Using it to compute the maximal Lyapunov exponent, I find a much faster
approach to the continuum limit, but a value of that limit which differs
by about $5 \%$ from the value in the Kogut-Susskind Hamiltonian system.
For the rate of $N_{CS}$ diffusion, the convergence to a fine lattice
limit is apprarently no faster, and the answers agree to within
error, but the statistical power of the test is limited, leaving open
the possibility that, as in the case of the maximal Lyapunov exponent,
there are lattice regulation artefacts which do not vanish in the $a=0$
limit.

Such an artefact presumably arises through 
interactions between the infrared modes, which determine the quanitites
of interest, and the ultraviolet excitations, which move under lattice
dispersion relations which differ between the two regulations, and
are in neither case rotationally invariant or ultrarelativistic.
This possibility arises because the ``hard thermal loop'' effects which the
ultraviolet modes induce are linearly divergent in the classical thermal
theory, so most of the contribution comes from the momentum region where
lattice effects are significant; and because the functional form the
hard thermal loop effects take is not strongly restricted by symmetries
of the theory, so the lattice dispersion relations of the ultraviolet modes
can cause the induced
hard thermal loops to have the wrong dependence on $\vec{q}/q_0$.  
Bodecker et. al.
have explicitly verified that this is the case for unimproved abelian 
Higgs theory \cite{McLerranetal}; in fact they show that the induced 
corrections to the infrared propagator are not even rotationally 
invariant.  There is every reason to believe that the same thing happens
in lattice Yang-Mills theory, in both the improved and unimproved cases.

What might at first seem surprising
is that the overall magnitude of the hard thermal loop effects, which 
in thermal units diverges as $1/a$, apparently does not affect the
infrared dynamical properties considered here, but the functional 
form of the hard thermal loops may.  This is also true
for one of the few infrared dynamical properties which is known for
the quantum theory, the damping rate of a plasmon at rest; 
in its calculation the 
presence of hard thermal loops is essential, and their dependence on $\vec{q}
/q_0$ determines the result, but their overall magnitude turns out to
cancel \cite{Pisarski}.  Something similar seems to be happening for the
maximal Lyapunov exponent.  Since neither Hamiltonian
considered to date produces the correct hard thermal loops, we cannot
take either limit to be correct.  The fact that they differ only by
a modest amount, and that both are correctly accounting for the thermodynamic
distribution of initial conditions, suggests that we can take the
answers determined from either Hamiltonian as a rough $(\sim 10 \%)$ 
estimate of the correct Lyapunov exponent and rate of $N_{CS}$ motion.
However, to improve on this accuracy in the case of the rate of $N_{CS}$
motion it will be necessary not only to determine with greater resolution
what the rate is in the Kogut-Susskind Hamiltonian system, but to check
with the same resolution that it is the same in the improved system.
If the answers do differ at the level of $5-10 \%$, then to get the
correct rate, some technique must be developed
which incorporates or generates hard thermal loops of the correct form.

\begin{center}
Acknowledgements
\end{center} 

I would like to thank Alex Krasnitz, for sparking my interest in this
matter and for helpful correspondence, and Chris Barnes, 
for useful conversations.  I would also like
to thank Neil Turok and Carl Edman, for donating computer resources.

All of the code used for the results presented here and in \cite{paper1}
is written in reasonably clear and well documented $c$, and is available
upon request to my electronic address, guymoore@puhep1.princeton.edu.

\pagebreak

\end{document}